\documentclass[%
 aip,
rsi,%
 amsmath,amssymb,
 reprint,%
]{revtex4-1}

\usepackage{graphicx}
\usepackage{dcolumn}
\usepackage{bm}
\usepackage[]{natbib}
\usepackage{color}

\usepackage[english]{babel}
\usepackage[utf8]{inputenc}

\usepackage[colorinlistoftodos]{todonotes}
\usepackage{siunitx}
\usepackage{setspace}
\usepackage{float}
\usepackage{xcolor}

\usepackage[export]{adjustbox}
\usepackage{parallel,enumitem}
\usepackage{array}
\usepackage{epstopdf}
\usepackage{eulervm}
\usepackage[toc,page]{appendix}
\usepackage{xspace}
\usepackage[T1]{fontenc}
\usepackage{hyperref}
\usepackage{silence}
\usepackage{lipsum}
 \hypersetup{
    colorlinks,
    citecolor=black,
    filecolor=black,
    linkcolor=black,
    urlcolor=black
}

\newcommand{\e}[1]{\vphantom{\mathbf{H}^{#1}}\mathbf{e}^{#1}}

 \newcommand{\sara}[1]{\textcolor{red}{#1}}

\begin{document}

\title{SAofSMCopolymers:Draft_1}


\title[]{Supramolecular copolymers predominated by alternating order: {{theory and application}}}
\author{Reinier van Buel}
\affiliation{Institute of Physics, Johannes Gutenberg-Universität, Staudingerweg 7-9, 55128 Mainz, Germany}
\affiliation{Theory of Polymer and Soft Matter Group, Department of Applied Physics, Eindhoven University of Technology, P.O. Box 513, 5600 MB Eindhoven, The Netherlands}

\author{Daniel Spitzer}
\affiliation{Institute of Organic Chemistry, Johannes Gutenberg-Universität Mainz, 55128 Mainz, Germany, Germany}

\author{Paul van der Schoot}
 \affiliation{Theory of Polymer and Soft Matter Group, Department of Applied Physics, Eindhoven University of Technology, P.O. Box 513, 5600 MB Eindhoven, The Netherlands}
 \affiliation{Institute for Theoretical Physics, Princetonplein 5, 3584 CC Utrecht, Utrecht University, The Netherlands}

\author{Pol Besenius}
\affiliation{Institute of Organic Chemistry, Johannes Gutenberg-Universität Mainz, 55128 Mainz, Germany, Germany}


\author{Sara Jabbari-Farouji}
\email{sjabbari@uni-mainz.de}
\affiliation{Institute of Physics, Johannes Gutenberg-Universität, Staudingerweg 7-9, 55128 Mainz, Germany}


\date{\today}

\keywords{Self-assembly, Supramolecular  copolymerization, Alternating copolymers,  Ising model } 
\begin{abstract}
We  investigate the  copolymerization  behavior of a  two-component system into quasi-linear self-assemblies  under  conditions that interspecies binding  is  favored over identical species binding. The theoretical framework is  based on a  coarse-grained  self-assembled Ising model with  nearest neighbor interactions. In Ising language, such conditions  correspond to the anti-ferromagnetic  case  giving rise to copolymers  
with predominantly alternating configurations.  
In the strong coupling  limit, we show that the maximum fraction of polymerized material and the average length of strictly alternating copolymers depend on the stoichiometric ratio and the activation free energy of the more abundant species. They are substantially reduced when the stoichiometric ratio noticeably differs  from unity.  Moreover, for stoichiometric ratios close to unity, the copolymerization critical concentration is remarkably lower than the homopolymerization critical concentration of either species.  We further analyze the polymerization behavior for a finite and negative coupling constant and  characterize the composition of supramolecular copolymers. Our theoretical insights rationalize experimental results of supramolecular polymerization of oppositely charged monomeric species in aqueous solutions. 


\end{abstract}

\maketitle

\section{Introduction} 
\label{sec:intro}

Biomolecular structures are typically formed of small building blocks that self-assemble into very complex and often symmetrical architectures.~\cite{whitesides2002,schneider2013,kirschner1986,lehn2002,cragg2010supramolecular}
These building blocks have a configurational functionality that allow for specific interactions between them. Their self-assembly thus leads to a high degree of control over the geometry of the assembled structures, which is critical for their functionality. For instance, the conformation of a protein and its biological functions are intrinsically linked.~\cite{lehn2002,rigden2009protein,buxbaum2007fundamentals,cragg2010supramolecular} 
Further examples of self-assembled structures in nature are phospholipid membranes, nucleic acids and ribosomes.~\cite{klug1983,kirschner1986}
In order to develop supramolecular polymers, a reversible self-assembly approach has been employed. In this method, the monomeric building blocks assemble through weak
non-covalent bonds, \textit{i.e.}, interactions on the order of the thermal energy~\cite{lehn1995,krieg2016,Sijbesma,Greef2009}, to yield larger polymeric structures. 
Typical supramolecular interactions are reversible coordination bonds, hydrogen 
bonds and electrostatic interactions, $\pi-\pi$ stacking, and van der Waals interactions. ~\cite{cragg2010supramolecular,Binder,PaulrevSMP,Brunsveld,Chakra,Rainer}

  Among naturally occurring self-assemblies, virus particles are prominent examples that have served as a source of inspiration for biomimetic supramolecular polymerization. 
A large part of the self-assembly into supramolecular structures and genome packaging into protective capsids is driven by electrostatic contributions. ~\cite{klug1983,Kushner,kegel2006,hagan2014,bruinsma2003,kegel2006,PaulrevSMP}  A viromimetic self-assembly strategy has further been applied to develop candidates for drug delivery vehicles, optoelectronic devices, sensors and medical diagnostics.~\cite{flynn2003viruses} 
The self-assembly of monomeric building blocks through electrostatic interactions is thus a powerful tool to create new nano-structured materials with tuneable functionalities such as  
optical, electrical, or magnetic properties.~\cite{faul2003}  For example, Tomba \textit{et al.}~\cite{tomba2010} used electrostatic interactions to create self-assembled supramolecular structures of rubrene on a gold substrate. They were able to show that a
combination of long range electrostatic repulsion and short-range attractive interactions drives the self-assembly into characteristic 1D patterns.

\begin{figure}[t]
\centering
\includegraphics[width=0.5\textwidth]{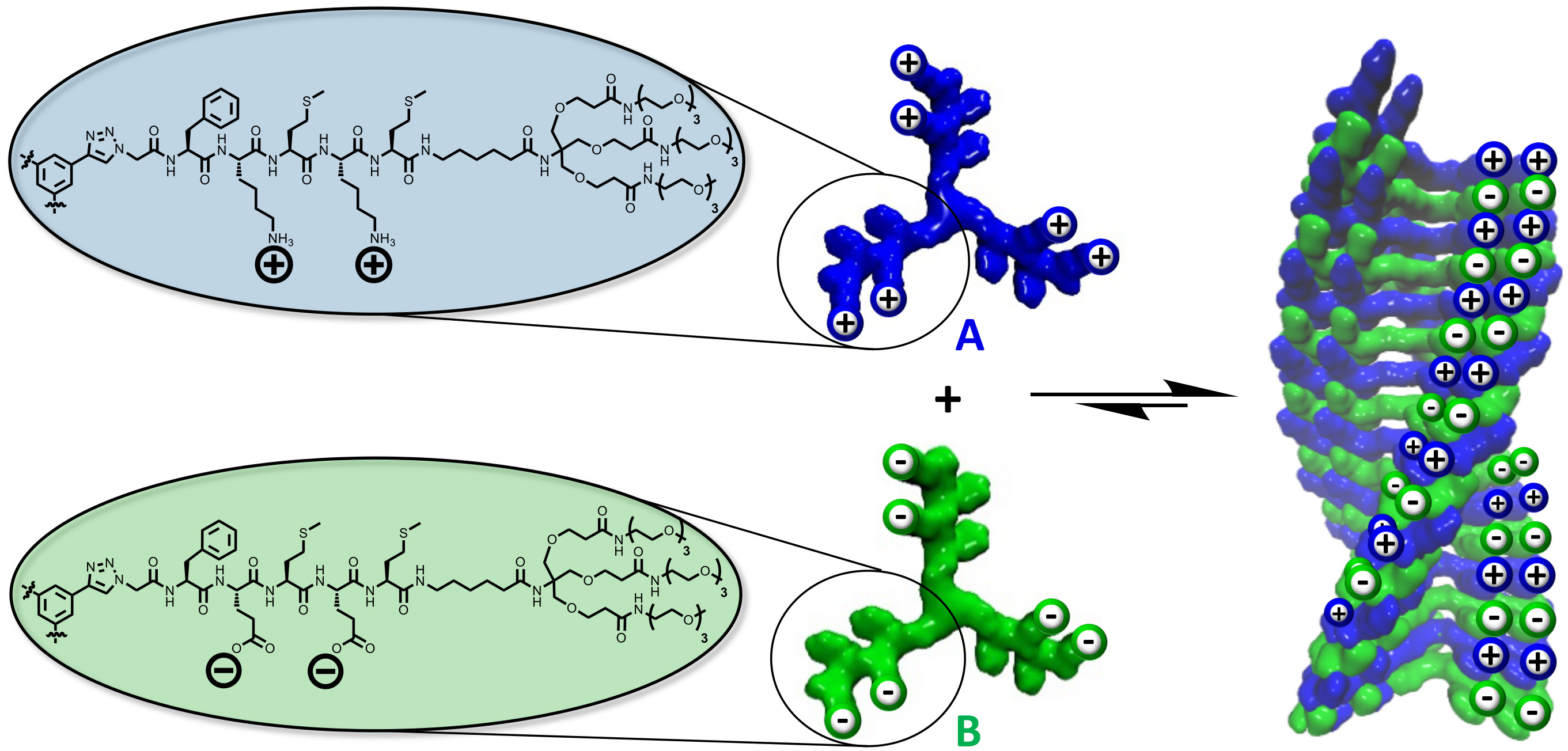}
  \caption{ Chemical structures of the cationic dendritic peptide monomer (A, blue) and the anionic complementary monomer (B, green), and a schematic representation of the self-assembly into alternating supramolecular copolymers.}
  \label{fig1}
\end{figure}

Recently, some of us~\cite{Frisch,Ahlers,Ahlers1} have developed a strategy to construct supramolecular copolymers using positively and negatively charged monomeric building blocks with $C_3$ symmetry, containing three identical peptide arms. 
 These arms contain amphiphilic oligopeptides, based on alternating sequences of hydrophobic phenylalanine or methionine and charged lysine or glutamic acid residues and form rod-like assemblies via a combination of electrostatic interactions, hydrogen bonding and hydrophobic shielding. ~\cite{Frisch}  
 The self-assembly is regulated via tuning the pH of the aqueous solutions.~\cite{Frisch,Ahlers1} 
 By changing the charged state of the monomeric building blocks, pH drives the self-assembly into homopolymers of a neutral monomer 
 or into copolymers when both  monomer species carry complementary charges.  A schematic
representation of the supramolecular copolymerization occurring at neutral pH $\approx7$ is presented in Fig. \ref{fig1}.
 
 
 Most recently, Ahlers \textit{et al.} performed a light scattering and electron microscopy investigation of the fidelity of
 the supramolecular copolymer formation.~\cite{Ahlers2}  
 The anionic and cationic peptide comonomers self-assemble into AB-type heterocopolymers, with a nanorod-like morphology and a thickness of 11 nm.
At equal concentrations of the two species,  the copolymers mean length  was  66 nm at an overall monomer concentration of $5 \times 10^{-5}$ M, equivalent to a volume fraction of $ \Phi \sim 3 \times 10^{-4}$.  
On the other hand,  excess in either  of the monomer species up to 50 mol \% in the stoichiometric ratio at the same overall  monomer concentration  decreased  the mean copolymer length to about $42-50$ nm.~\cite{Ahlers2} 
These findings prompted us to  gain a microscopic insight into the thermodynamic effects that dictate the configuration and formation of the copolymers
   from a theoretical perspective. Since circular dichroism  and fluorescence spectroscopy tools have so far not been able to directly determine the composition of the supramolecular copolymers  (alternating, random or blocky structures), we gain a deeper understanding of the morphology and optimal stoichiometric ratio for alternating copolymers
   using a two-component coarse-grained model.~\cite{Sara} 

 Coarse-grained models  for self-assembly  of monodisperse  systems ~\cite{cates1990,ciferri2005,porte1983,PaulrevSMP} 
 have played an essential role in rationalizing experimental data. These models successfully describe the equilibrium self-assembly behavior of 
 quasi-linear supramolecular homopolymers  for  the so-called isodesmic, activated and auto-catalytic assembly cases. 
There has been an increasing interest in a theoretical understanding of the self-assmebly behavior in multicomponent systems  and  self-assembly theories have been extended to  supramolecular copolymers consisting of two or more chemically distinct monomer species ~\cite{chelating,Gestel2003,diblock1,diblock2,diblock3,diblock4,Markvoort,Smulders2,Sara,SupramolecularCopolymers} to capture the physics of system under consideration.  
The 1D self-assembly theory developed in the reference~\cite{Sara} is relevant for the experiments described above.  It builds on a two-component self-assembled Ising model where the presence of two different species is parameterized in terms of the strengths of
binding free energies that depend on the monomer species involved in the pairing interaction. The theory predicts formation of different
morphologies of copolymer assemblies
depending on the relative values of the species-dependent binding free energies, exhibiting  random, blocky and alternating ordering of 
the two components in the assemblies.

The decisive parameter determining the arrangement of monomer species in the assemblies ({\it i.e.}, the morphology) is an effective \emph{coupling constant} of the Ising model
$ J \equiv \left(b_{AA}- 2\, b_{AB}+ b_{BB}\right)/4$, which depends on a linear combination of the binding free energies $b_{ij}$ between two monomer species
$i$,  $j\in\{A$,  $B\}$.  
The $J>0$ case, with favorable binding  between monomers of the same species, leads to formation of copolymers with blocky order. For the $J=0$ case 
monomers of the two species are
randomly distributed along the assemblies, whereas the $J< 0$  case leads to copolymers with predominantly alternating order. 
This model was thoroughly analyzed for the case of blocky ordering $J>0$.~\cite{Sara} Here, we analyze the model for the $J<0$ (antiferromagnetic coupling) case where the binding  between two distinct species is  favored over binding between
  those of the same type. We show that  our model can rationalize the experimental results for the copolymerization  of  the two-component system with complementary charges,  although it only incorporates nearest-neighbor interactions.
This approximation works well because the long range electrostatic interactions of these 1D co-assemblies are effectively reduced to a nearest neighbor contribution, due to \textit{self-screening}
effects in an alternating charge sequence.~\cite{Netz,Reinierthesis}  

The remainder of this article is organized as follows. In section \ref{sec:theory}, we briefly review the theoretical framework of
two-component supramolecular copolymers  and discuss 
its main ingredients and the resulting mass-balance equations. In section \ref{sec:Strongcoupling}, we analyze the mass-balance equations  in the strongly negative coupling limit $J \ll -1$, where interspecies binding is  predominant and obtain the dependence of the critical concentration on the stoichiometric ratio and the free energy parameters. We investigate the copolymerization behavior  for the case of finite and negative $J$ in section \ref{sec:results}. Finally, we compare our theoretical insights to experimental findings in section \ref{sec:exp} and our main conclusions can be found in section \ref{sec:conclusion}.


\section{Review of theory of Supramolecular co-polymerization }
\label{sec:theory}

In this section, we briefly review  the supramolecular polymerization theory of linear assemblies for  two-component systems developed in reference.~\cite{Sara} We consider   two  monomeric species, $A$ and $B$,  with equal effective interaction volumes: $\nu_A=\nu_B=\nu$.
The conformational free energy difference between a free monomer of type $i\in\{A,B\}$ in the solution and  one 
bound to an assembly is parametrized  by an activation free energy $a_{i}>0$. The free energy gain of bonded interactions between  two species $i$ and $j$  is parametrized by $-b_{ij} $, where $i$, $j\in\{A,B\}$. All the free energies are scaled to the thermal energy $k_B T\equiv 1$.
The semigrand potential energy of a dilute solution of  volume $V$ containing free monomers and self-assembled polymers  can be expressed by  a sum of an ideal mixing  entropy and the contributions from the internal partition  functions of  assemblies with varying degrees of polymerization $N$:
\begin{equation}
\!\!\frac{\Omega}{V}=\!\!\sum_{N=1}^\infty \!\! \rho(N)\left[\ln \rho(N)\nu-1-\ln Z_N(\mu_i,b_{ij},a_{i}) \right].\!\!
\end{equation}
Here,  $\rho(N)$ is the number density of assemblies containing $N$ monomers and $\mu_i$ represents the chemical potential of species $i\in\{A,B\}$. The semi-grand canonical partition function $Z_N(\mu_i, b_{ij}, a_{i})$ 
accounts for the Boltzmann weighted sum of all the conformational states of assemblies with size $N$.

Minimizing the grand potential energy $\delta \Omega/\delta \rho(N)=0$ yields  the equilibrium size distribution of assemblies: 
\begin{align}
\nonumber
\rho(1)\nu =& \;  Z_1=\mathbf{e}^{\mu_A+a_A} + \mathbf{e}^{\mu_B+a_B},
\\
\rho(N)\nu =& \; Z_N \hspace{2cm}  \; (N > 1).
\end{align}
The partition function $Z_{N}$  of assemblies in  a two-component system can be
obtained by mapping the Hamiltonian of  linear assemblies onto an Ising model  with an effective Hamiltonian of the form ~\cite{Sara}:
\begin{align}
\nonumber
-\mathcal{H}_{N>1}  = 
\, J \sum_{l=1}^{N-1} &  S_l S_{l+1} + H \sum_{l=1}^N S_l + E_0(N) 
\\
\label{Hamil}
& -(S_1 + S_N) (b_{AA}-b_{BB})/4,
\end{align}
\noindent where $S_l=\pm 1$ is the spin state of the site $l$ that is $1$ when the site is occupied by species $A$ and $-1$ when  occupied by species $B$.
The effective coupling constant $J$ between two neighboring sites is given by
\begin{equation}
\label{eq:J}
 J\equiv \left(b_{AA}- 2\, b_{AB}+ b_{BB}\right)/4, 
\end{equation}
and
\begin{equation}
H \equiv \left( b_{AA} - b_{BB} + \mu_A - \mu_B \right)/2, 
\end{equation}
describes the effective external field that couples to the spin sites. It is directly linked to the   difference in the chemical potential of the two species $\Delta \mu\equiv (\mu_A-\mu_B)/2$.
Furthermore, the spin-independent term
\begin{align}
E_0(N)\equiv (N-1) \; \bar{b} + N \; \bar{\mu}
\end{align}
defines the average  binding free energy  $\bar{b} = (b_{AA}+b_{BB}+2b_{AB})/4$ and the average chemical potential   $\bar{\mu} = (\mu_A + \mu_B )/2$ of the two-component system.

The coupling parameter $J$ is the key quantity that determines the composition  of monomers in the self-assembled polymers. 
For $J>0$ ferromagnetic ordering is favored, implying blocky copolymers. $J=0$ corresponds to paramagnetic ordering meaning random copolymers. 
The $J<0$ case leads to anti-ferromagnetic  ordering associated with alternating order in  copolymers. It  is of 
  relevance to the experiments of oppositely charged monomers and will be thoroughly analyzed in the following sections.  

Using the standard transfer matrix method ~\cite{goldenfeld1992}, the  partition function of an assembly of size $N>1$ with open boundary conditions becomes~\cite{Sara}
\begin{equation}
\label{eqpartition}
Z_{N>1} = \left( x_+ \lambda_+^{N-1} + x_-\lambda_-^{N-1} \right) \exp E_0 (N),
\end{equation}
where the eigenvalues of the transfer matrix $\lambda_\pm$ are given by 
\begin{align}
\label{eqev}
\lambda_\pm =\e{J} \cosh (H) \pm \e{-J} \sqrt{\e{4 J} \sinh ^2(H)+1}.
\end{align}
 The coefficients  $x_\pm$ are given by ~\cite{Sara}
\begin{align}
\label{eqx}
x_\pm =\cosh(\Delta \mu) \pm \frac{1 + e^{2 J} \sinh(\Delta \mu) \sinh (H)}{ \sqrt{e^{4 J}\sinh ^2(H)+1}},
\end{align}
 describing the statistical weight associated with different possible configurations at the chain ends. They arise 
  from the open boundary condition that implies either of species $A$ or $B$ can be located on either end of a chain.

 To determine the values of the chemical potentials, we impose mass conservation  for both species. Denoting
 the  volume fraction of the  species $i$  by $\Phi_i$, the mass balance equation for the overall volume fraction $\Phi \equiv \Phi_A + \Phi_B$ obeys
\begin{eqnarray}
\label{eq:Totalvolfrac}
\Phi &=& \sum_{N=1}^\infty N \rho(N) \nu   \\ \nonumber
&=& \mathbf{e}^{\mu_A}(\mathbf{e}^{a_A}\!-\!1) + \mathbf{e}^{\mu_B}(\mathbf{e}^{a_B}\!-\!1)+\sum_{\sigma=\pm} \frac{ x_\sigma \mathbf{e}^ {\bar{\mu}}} {(1-\Lambda_\sigma)^2},
\end{eqnarray}
where   $\Lambda_\pm = \lambda_\pm \exp(\bar{\mu}+\bar{b})$ represent effective fugacities of the bidisperse system.
 
Similarly, the difference in  the  volume fraction of the two species $\Delta \Phi\equiv \Phi_A - \Phi_B$ can be expressed in terms of the average spin value $\langle S(N) \rangle$ as 
$\Delta \Phi = \sum_{N=1}^\infty N \rho(N) \nu \langle S(N) \rangle$, given by
 %
\begin{eqnarray}
\label{eq:Diffvolfrac}
\Delta \Phi 
 &=&  \mathbf{e}^{\mu_A}(\mathbf{e}^{a_A}\!-\!1)- \mathbf{e}^{\mu_B}(\mathbf{e}^{a_B}\!-\!1) \\ \nonumber
 &+&  \sum_{\sigma=\pm} \frac{ x_\sigma \Lambda_\sigma \mathbf{e}^{\bar{\mu}}} {(1-\Lambda_\sigma)^2} \big( (1-\Lambda_\sigma) \frac{\partial \ln (x_{\sigma})}{\partial H}+ \frac{\partial \ln (\lambda_{\sigma})}{\partial H}\big).
\end{eqnarray}
The coupled set of equations \eqref{eq:Totalvolfrac} and  \eqref{eq:Diffvolfrac} are the central equations that will be solved throughout the rest of this paper to obtain the chemical potentials $\mu_i$.

By determining the chemical potentials from the above set of equations, the mean degree of polymerization 
\begin{equation}
\label{eq:meandegree}
\overline{N} \equiv \frac{\sum_{N=1}^\infty N \rho(N)}{\vphantom{\sum^2}\sum_{N=1}^\infty \rho(N)}
= \frac{\Phi}{\vphantom{\sum^2}\sum_{N=1}^\infty \rho(N)\nu},
\end{equation}
is straightforwardly evaluated because the sum in the denominator can be simplified to
\begin{equation}
\sum_{N=1}^\infty \rho(N)\nu = \sum_{\sigma=\pm} \frac{ \mathbf{e}^{\bar{\mu}} x_\sigma\Lambda_\sigma}{ (1-\Lambda_\sigma)}+  \mathbf{e}^{\mu_A+a_A} + \mathbf{e}^{\mu_B+a_B}.
\end{equation}
Moreover, the fraction of polymerized material obeys the simple relation
\begin{equation}
f =  1 - \frac{\rho(1)\nu}{\Phi} = 1 - \frac{\mathbf{e}^{\mu_A+a_A} + \mathbf{e}^{\mu_B+a_B}}{\Phi},
\end{equation}
which is, in principle, an experimentally accessible quantity.
As expected, in the  limit  of vanishing  concentration of  $j$ species   ($\mathbf{e}^{\mu_j} \to 0$),  the theoretical model recovers all governing equations of
a homopolymer system made of the other species  $i\neq j$.
In this limit, the effective fugacity obeys $\Lambda_+= \exp(\mu_i+ b_{ii})$ and the mass balance equation reduces to
\begin{align}
\label{eq:mono}
\frac{\Phi}{\Phi^*}=\Lambda_+  \mathbf{e}^{-a_i} \Lambda_+^2 \frac{2-\Lambda_+}{(1-\Lambda_+)^2},
\end{align} 
in which $\Phi^* = \exp(a_{i}-b_{ii})$ is the so-called critical concentration.~\cite{Paulreview}  It demarcates the transition from a monomer-dominated to a polymer-dominated regime.

The analytical solution of the coupled mass balance equations  (\ref{eq:Totalvolfrac}-\ref{eq:Diffvolfrac})  is not generally known. However, for the special case of 
$b_{AA}=b_{BB}$, $a_A=a_B=a$ and $\alpha=1$, we can solve the mass-balance equations. In this case, Eq.~\eqref{eq:Diffvolfrac} yields $z_A=z_B$ and Eq.~\eqref{eq:Totalvolfrac}  becomes identical to the monodisperse mass balance equation, Eq.~\eqref{eq:mono}, in which $\Lambda_+\equiv 2 z_i \exp b_{eff}$ and $\Phi^* \equiv  \exp(-b_{\text{eff}}+a)$. Here, $b_{\text{eff}}\equiv \bar{b}+\ln[ \cosh (J)]$ is an effective binding free energy that incorporates the effect of mixing of the two components in the assemblies.  In the following, we analyze the copolymerization behavior for the case of negative coupling constant  $J<0$ .

\section{Strongly negative coupling limit: strictly alternating copolymers}
\label{sec:Strongcoupling}

 In this section,  we examine the situation that the free energy gain of interspecies binding  is much greater than that of identical species binding, {\it i.e.}, $b_{AB}\gg b_{AA}$ and $b_{AB}\gg b_{BB}$. We consider two distinct cases: i) $b_{ii} \ll -1$, {\it i.e.},  no binding takes place between identical species and $b_{AB} >0$ can have any finite value and ii)  $b_{AB} \gg 1$ and $b_{ii}$ have finite values.
As can be seen from Eq.~\eqref{eq:J}, either of these conditions lead to $J \ll -1$ for which  comonomers along the assemblies predominantly presume an alternating order. 
We first simplify the mass balance equations, Eqs.~(\ref{eq:Totalvolfrac}-\ref{eq:Diffvolfrac}), in the limit of $J \rightarrow -\infty$ and analyze the copolymerization behavior.  Next, we inspect the dependence of the  critical concentration on the stoichiometric ratio.

\subsection{Mass balance equations}

In the limit $e^{J} \ll 1$,  the eigenvalues $\lambda_\pm$ in Eq.~\eqref{eqev}  simplify to
\begin{align}
\label{eqEVtaylor}
\lambda_\pm &= \pm \mathbf{e}^{-J}+ \cosh(H) e^{J}+ \mathcal{O}(e^{2J}).
\end{align}

As a result, the effective fugacities reduce to
\begin{align}
\Lambda_\pm &= \pm  \mathbf{e}^{\bar{\mu}+b_{AB}} +  \mathbf{e}^{\bar{\mu}+(b_{AA}+b_{BB})/2}\cosh(H) +\mathcal{O}(e^{2J}).
\label{eqL1st}
\end{align}
This equation shows that in such a strongly negative coupling  limit,  the leading order term of the copolymers effective fugacities depend only on
  $b_{AB}$  and $\bar{\mu}$. The second leading order term is independent of $b_{AB}$ and only depends on $b_{ii}$  and $ \mu_{i}$.

  Likewise, Taylor expanding Eq.~\eqref{eqx} up to  the first order in $ e^J$,  the end cap weights $x_\pm$ in the strongl coupling  regime reduce to
\begin{align}
\label{eqx1st}
x_\pm &= \cosh(\Delta \mu) \pm 1 + \mathcal{O}(e^{2J}).
\end{align}
We note that in the strong coupling limit, the  end cap weights are symmetric with respect to the particle species.  

 Keeping only the leading order terms,  we obtain the $N$-dependent partition functions in the  $J \rightarrow -\infty$ limit.  
They can be split into  partion functions for assemblies  with even and odd degrees of polymerization given by
\begin{eqnarray}
\label{eqZeven}
Z_{N}^\text{even} &=& 2 \exp[(N-1)b_{AB}+ N \bar{\mu}](1+\mathcal{O}(e^{2J})) \nonumber \\
&=& 2 \exp[-\mathcal{H}_{AB}^{\text{even}}],
\end{eqnarray}
\begin{eqnarray}
\label{eqZodd}
Z_{N>1}^\text{odd} &=& 2\cosh(\Delta \mu) \exp[(N-1)b_{AB}+ N \bar{\mu}](1+\mathcal{O}(e^{2J}))  \nonumber \\
 &=&\exp[-\mathcal{H}_{AB}^{\text{odd}}]+ \exp[-\mathcal{H}_{BA}^{\text{odd}}].
\end{eqnarray}
Here, $\mathcal{H}_{AB}^{\text{even}}$ is the free energy of  an alternating copolymer consisting 
of $N/2$ units of $AB$ or $BA$ dimers. Likewise, 
   $\mathcal{H}_{AB}^{\text{odd}}$ and $\mathcal{H}_{BA}^{\text{odd}}$
  represent the free energies  of  alternating copolymers of the form  $B(AB)_{(N-1)/2} $ and $A(BA)_{(N-1)/2}$, {\it i.e.}, with an excess of $B$  or  $A$  monomers, respectively. Thus, in the limit $J \rightarrow -\infty$ the partition functions only depend on $b_{AB}$ and the chemical potentials and do not exhibit any singularity. In other words, in this limit our results converge to that of a bidisperse system where only interspecies binding takes place, {\it i.e.}, $b_{ii} \to -\infty$ and $b_{AB}$ can have  any arbitrary positive value.
    Notably, for the case of equal chemical potentials, the prefactor of partition function for assemblies with odd number of monomers, $\cosh(\Delta \mu) $ reaches its minimum value and $Z_{N}^\text{even}=Z_{N}^\text{odd}$. This implies that  the free energy cost of adding another monomer to a polymer with an arbitrary size is  $-(b_{AB}+\bar{\mu})$ and independent of $N$. For unequal chemical potentials the fraction of
alternating polymers with even degree of polymerization is smaller than those with odd degree of polymerization 
because the more abundant species, assuming to be $B$,   can have a greater contribution to the polymerization by
forming  $B(AB)_{(N-1)/2} $  copolymers.


Using the leading order terms of Eqs. (\ref{eqEVtaylor}-\ref{eqx1st}), the  mass balance equation for the overall volume fraction of monomers,  given by Eq.~\eqref{eq:Totalvolfrac},  simplifies to
\begin{eqnarray}
\nonumber
\Phi &=& \frac{(\mathbf{e}^{\mu_A}+\mathbf{e}^{\mu_B})\left(1+\Lambda_+^2\right)+4 \mathbf{e}^{-b_{AB}} \Lambda_+^2}{\vphantom{\sum^2}(1-\Lambda_+^2)^2}
\\
& + & \mathbf{e}^{\mu_A} (\mathbf{e}^{a_A} -1) + \mathbf{e}^{\mu_B} (\mathbf{e}^{a_B} -1),  
\label{eq:totvolfraclim}
\end{eqnarray}
in which  $\Lambda_+ \equiv \exp (\bar{\mu}+b_{AB}) $.
Likewise, using $\partial/ \partial{H}=  \partial/ \partial{\Delta \mu} $, Eq.~\eqref{eq:Diffvolfrac}  expressed in terms of the stoichiometric ratio $\alpha \equiv \Phi_A/\Phi_B$ simplifies to 
\begin{eqnarray} \label{eq:diffvolfraclim}
\nonumber
\Phi \frac{\alpha-1}{\alpha+1} &=&   \frac{(\mathbf{e}^{\mu_A}- \mathbf{e}^{\mu_B})\Lambda_+^2}{1-\Lambda_+^2}  \\  
&+&   \mathbf{e}^{\mu_A+a_A}  -  \mathbf{e}^{\mu_B+a_B} .
\end{eqnarray}
The first term in the right hand side of the above equation arises from the asymmetry in the population of copolymers with odd degrees of polymerization. 
For stoichiometric ratios different from unity, odd-numbered copolymers that include a larger fraction of  the more abundant species are preferred. 

 Even in the strong  coupling limit, we can not analytically  solve  this set of coupled equations for  the chemical potentials.
Only for the special case  of equal concentration of the two species,   $\alpha=1$, and  identical activation energies \mbox{$a_A=a_B=a$}, we can obtain an analytical solution for the mass balance equations. Under such conditions, Eq.~\eqref{eq:diffvolfraclim} yields $\mu_A=\mu_B$, {\it i.e.},  the concentration of free monomers of the two species are equal at any $\Phi$.  The mass balance equation   Eq.~\eqref{eq:totvolfraclim} for the overall concentration simplifies to  $\Phi=2 \mathbf{e}^{\mu_A}[(1-\Lambda_+)^{-2}+(\mathbf{e}^{a_A}-1)]$. It can be expressed as a third order polynomial in terms of $\Lambda_+ \equiv \exp (\bar{\mu}+b_{AB}) $ given by
\begin{equation}
\label{eq:AB-polymer}
 \frac{\Phi}{\Phi^*_{AB}}= \Lambda_+ + \mathbf{e}^{-a} \Lambda_+^2 \frac{2-\Lambda_+}{(1-\Lambda_+)^2},
\end{equation}
%
where we have introduced the critical concentration $\Phi^*_{AB} \equiv 2 \e{-b_{AB}+a}$. Notably, this equation has exactly the same form as the mass balance equation of a monodisperse system given in Eq.~\eqref{eq:mono}.
Therefore, alternating copolymers can be thought of as   homopolymers composed of  $(AB)$-dimer building units. 
An identical form for the mass balance equations in the two cases suggests that  $\Phi^*_{AB}$ can be interpreted as  the critical concentration of alternating copolymers. In subsection \ref{sect_critC}, we obtain the   dependence of the  critical concentration on $\alpha$ for sufficiently negative $J$
and  show that  $\Phi^*_{AB}$ is indeed the  critical concentration for the $\alpha=1$ case, when $b_{ii} \rightarrow -\infty$. Before that we first
investigate the solution of the simplified mass-balance equations.

\subsection{Polymerization behavior} \label{sec:strict_alt}
\begin{figure}[t]
\centering
  \includegraphics[width=.95\linewidth]{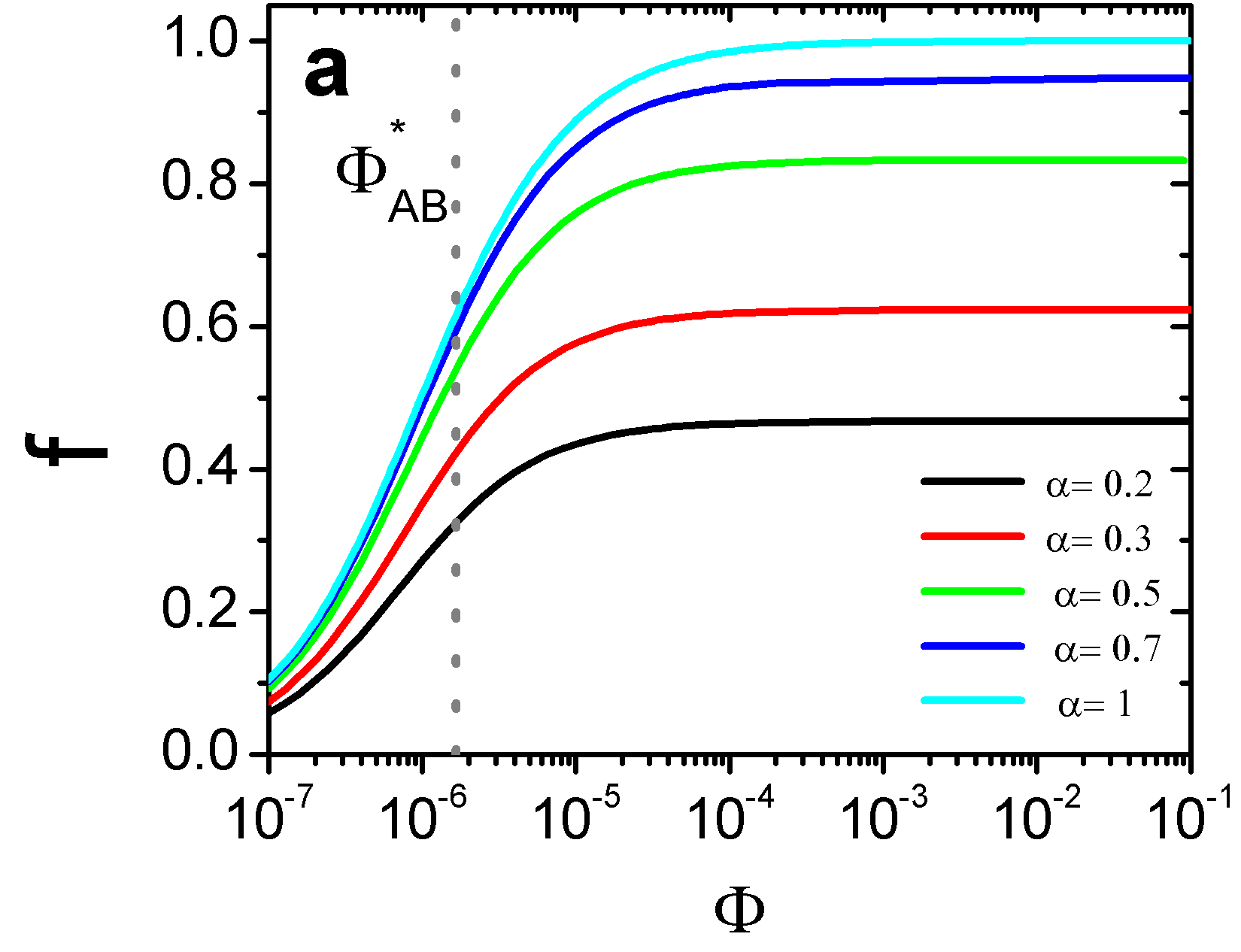}
  \includegraphics[width=.95\linewidth]{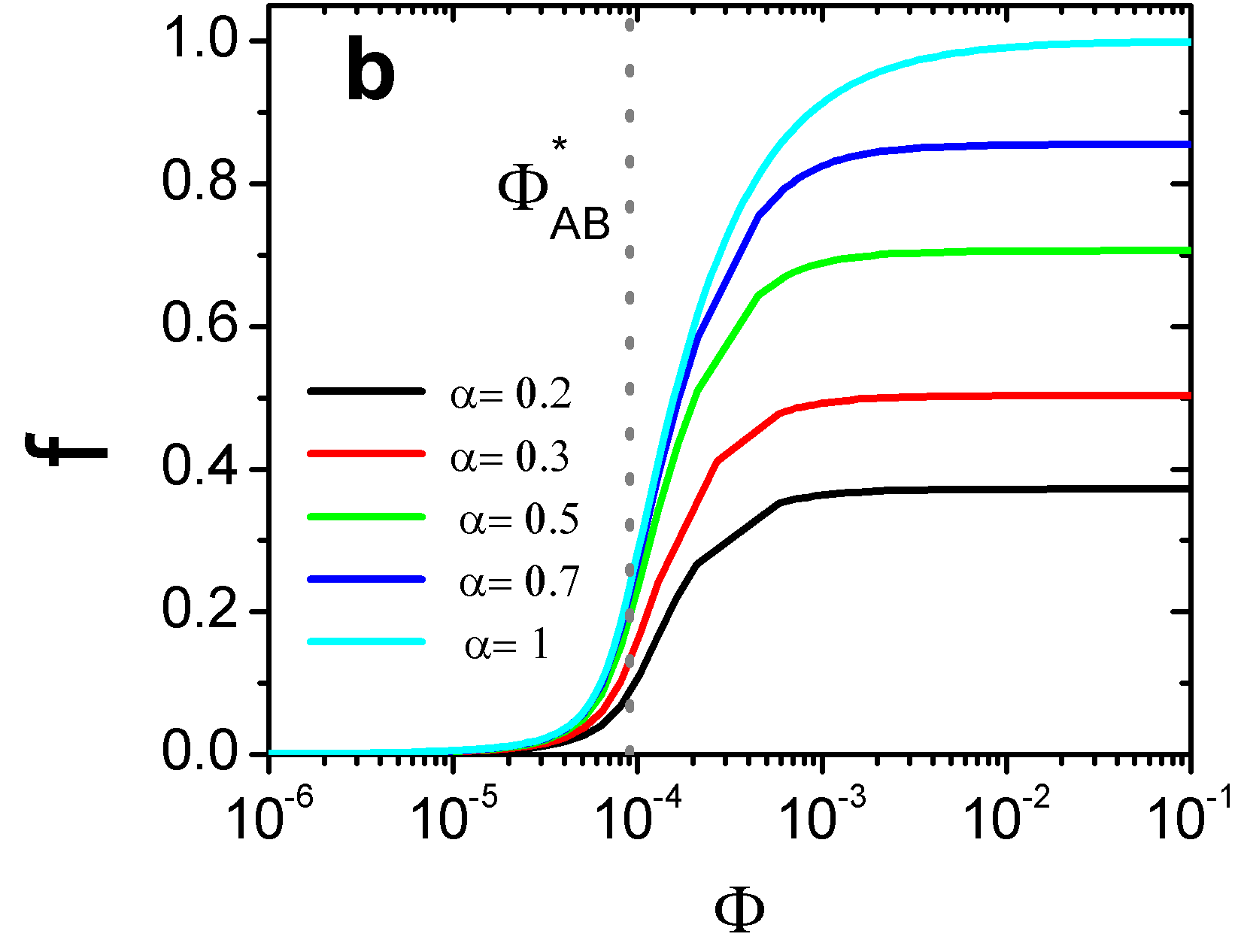}
  \caption{Fraction of polymerized material $f$ as a function of the overall volume fraction  $\Phi=\Phi_A+\Phi_B$ at different stoichiometric ratios $\alpha=\Phi_A/\Phi_B$ in the limit of $J \rightarrow - \infty$ where no binding between identical species  takes place, { \it i.e.}, $b_{AA}=b_{BB} \rightarrow -\infty$. The assumed binding free energy between the two distinct species is  $b_{AB}=b_{BA}=14$ and the activation free energies in panels (a) and (b) are $a_A=a_B=0$ and   $a_A=a_B=4$, respectively. The dotted lines show the critical concentration  $\Phi^*_{AB}=1.66 \times 10^{-6}$ and $\Phi^*_{AB}= 9.08 \times 10^{-5}$ for $a=0$  and $a=4$, respectively. }
  
  \label{fig:f_alt}
\end{figure}
In this subsection, we analyze the polymerization behavior in the strongly negative coupling limit, $J \rightarrow -\infty$,  for the case that $b_{AA}=b_{BB} \rightarrow -\infty$, leading to formation of strictly alternating copolymers. As pointed out in the previous subsection, 
 the mass balance  equations in this limit do not exhibit any singularity and are valid for  arbitrary values of $b_{AB}$. The only  variables appearing in the resulting mass balance equations are the interspecies binding energy $b_{AB}$ and activation free energies $a_i$. 
We fix the value of the binding free energy between the two comonomers to $b_{AB}=14$. 
To reduce the number of parameters  in the mass balance  equations, Eq. \eqref{eq:totvolfraclim} and Eq.\eqref{eq:diffvolfraclim},  we assume equal activation free energies $a_A=a_B=a$. 
We investigate the polymerization behavior of copolymers with strictly alternating order as a function of the overall concentration
and the stoichiometric ratio for different values of activation free energy. We restrict our analysis to  $0<\alpha<1$, as the results for the  $\alpha >1$ case can be obtained by  an $\alpha \rightarrow 1/\alpha$ transformation.

Figs.~\ref{fig:f_alt}a and \ref{fig:f_alt}b show  the fraction of polymerized material $f$  as a function of the total volume fraction $\Phi$ at
several values of the stoichiometric ratio, for $a=0$ and $a=4$, respectively. In all the cases, we observe a transition from a monomer-dominated  to
a polymer-dominated regime for which the fraction of polymerized material reaches a maximum value, $f^{\text{max}}$, that depends on the stoichiometric ratio and the activation free energy. Notably,  the $f^{\text{max}}$ of  copolymers with strictly alternating order is smaller than one except for the case of perfect stoichiometric balance $\alpha=1$.  A  $f^{\text{max}}$ smaller than one reflects the lack of the less abundant species, {\it i.e.}, type $A$ monomers for the  $\alpha < 1$ case. Copolymers with odd degrees of polymerization mainly exist in the $B(AB)_{(N-1)/2}$ form  rather than the $A(BA)_{(N-1)/2}$ form  to consume a larger amount of the $B$ species. Similar to homopolymerization transition~\cite{Paulreview}, the sharpness of copolymerization transition depends on the  strength of the activation free energy. It becomes steeper for larger  activation free energies $a$.  We can characterize the transition point as the  monomeric volume fraction for which $f \approx 2/3 f^{\text{max}}$. This value shifts to  a higher value for the larger activation free energy and it roughly agrees with the critical concentration value defined earlier as $\Phi^*_{AB} \equiv 2 \e{-b_{AB}+a}$.  The dotted lines in Figs.~\ref{fig:f_alt}a and \ref{fig:f_alt}b correspond to  $\Phi^*_{AB}=1.66 \times 10^{-6}$ and $\Phi^*_{AB}= 9.08 \times 10^{-5}$ for $a=0$  and $a=4$, respectively. 

\begin{figure}[t]
\centering 
  \includegraphics[width=.95\linewidth]{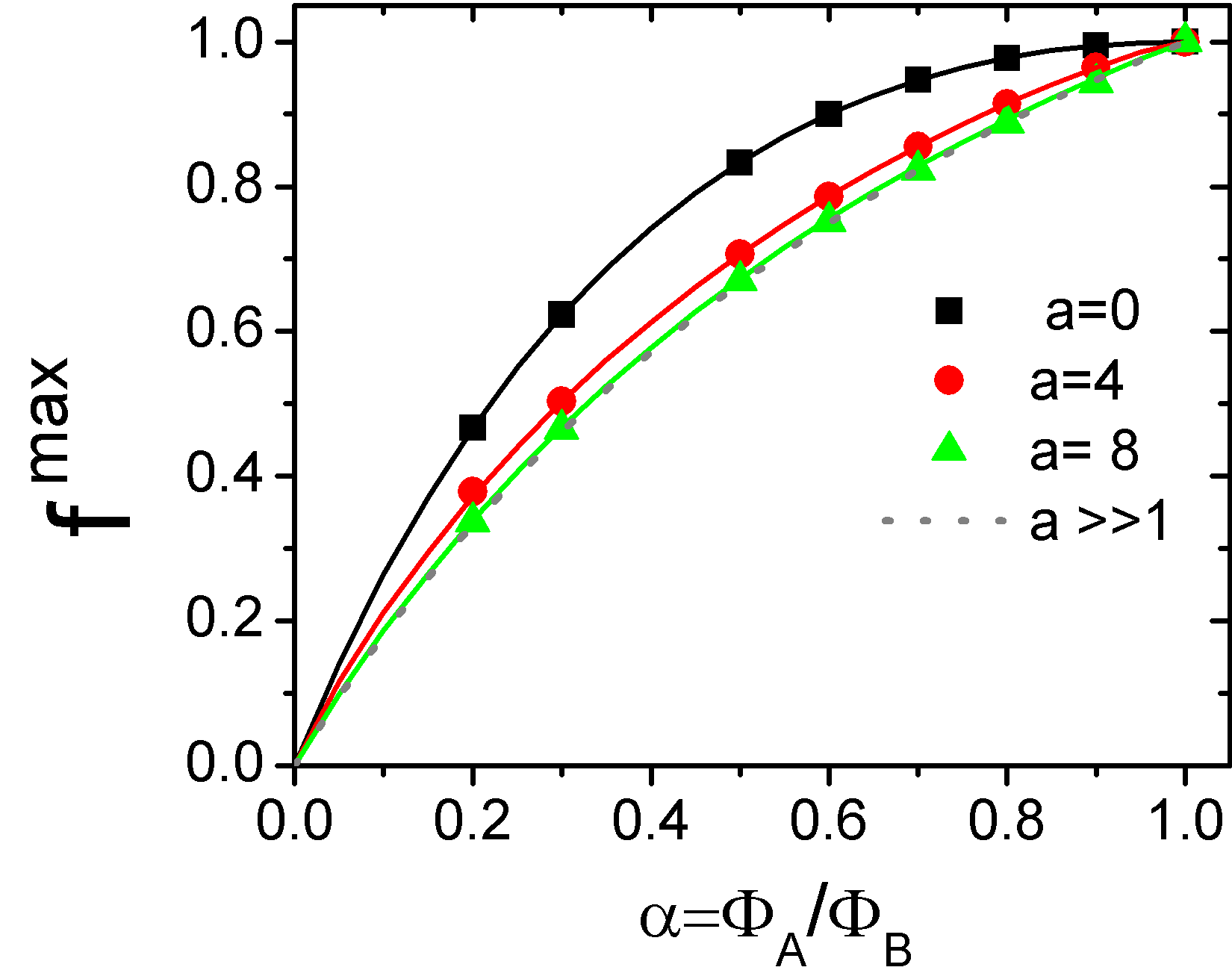}
     \caption{ The maximum fraction of polymerized material in the saturation limit, $f^{\text{max}}$, as a function of  the stoichiometric ratio  $\alpha=\Phi_A/\Phi_B <1$ for different values of activation free energies $ a_A=a_B=a $,  in the limit of $J \rightarrow - \infty$, where no binding between identical species occurs, { \it i.e.}, $b_{AA}=b_{BB} \rightarrow -\infty$.  The symbols show $f^{\text{max}}$ values obtained from the numerical solution of the mass balance equations Eqs.~(\ref{eq:totvolfraclim}- \ref{eq:diffvolfraclim}) for interstices binding free energy $b_{AB}=b_{BA}=14$ at high concentrations. The lines show our theoretical predictions given by Eq.~ \eqref{eq:fmax} and   Eq.~\eqref{eq:fmaxlimit} confirming that $f^{\text{max}}$ is independent of $b_{AB}$. }  
  \label{fig:f_max}
\end{figure}

   At large $\Phi$, in the saturation regime where most of the type $A$ monomers are consumed, we expect $\mu_A \ll \mu_B$. 
   We can use this approximation to solve the mass balance equations Eq.~\eqref{eq:totvolfraclim} and Eq.~\eqref{eq:diffvolfraclim}   analytically and estimate the fraction of polymerized material as $f^{\text{max}} \approx 1-\mathbf{e}^{\mu_B+a}/ \Phi$. Doing so, we obtain the functional dependence of $f^{\text{max}}$ on $a$ and $\alpha$ as
\begin{equation}
 f^{\text{max}}=\frac{4 \alpha  \left(e^{a}-1\right)+\sqrt{1+4 (1-\alpha ) \alpha  \left(e^{a}-1\right)}-1}{2 (\alpha +1). \left(e^{a}-1\right)},
  \label{eq:fmax}
 \end{equation}
 that becomes independent of the binding free energies. 
 We again note that these results are valid for $0<\alpha <1$ and the results for $ \alpha > 1$ can be obtained by replacing $\alpha$ with $1/\alpha$.  Moreover, in the case of unequal activation free energies, $a$ represents the activation free energy of the more abundant species.
 $f^{\text{max}}$ can be simplified further in  two limiting cases of $a \rightarrow 0$ and $a \to \infty$ to
\begin{equation}
    f^{\text{max}} =
    \begin{cases}
      \frac{  \alpha (3-\alpha) }{  (\alpha +1) } & a  \rightarrow 0 \\
              \frac{   (2 \alpha) }{  (\alpha +1) } & a \to \infty .
       \end{cases}
       \label{eq:fmaxlimit}
  \end{equation}
 In the extreme case of $a \rightarrow \infty$ and $b_{AB} \rightarrow \infty$ such that  $-b_{AB}+a$ is finite, the polymerization transition corresponds to a sharp transition from a monomeric regime to a copolymer dominated regime.  In this case,  equal volume fractions of A and B  monomers ($\alpha/(1+\alpha) \Phi$)  are self-assembled and the excess of $B$ monomers remains in the solution.    The $f^{\text{max}}$ values extracted from the numerical solution of the mass balance equations are presented in Fig. \ref{fig:f_max}. They show
  excellent agreement with our theoretical predictions given by Eq.~\eqref{eq:fmax} and Eq.~\eqref{eq:fmaxlimit} for all values  of $a$ and $\alpha$. 

\begin{figure}[t]
\centering 
  \includegraphics[width=.95\linewidth]{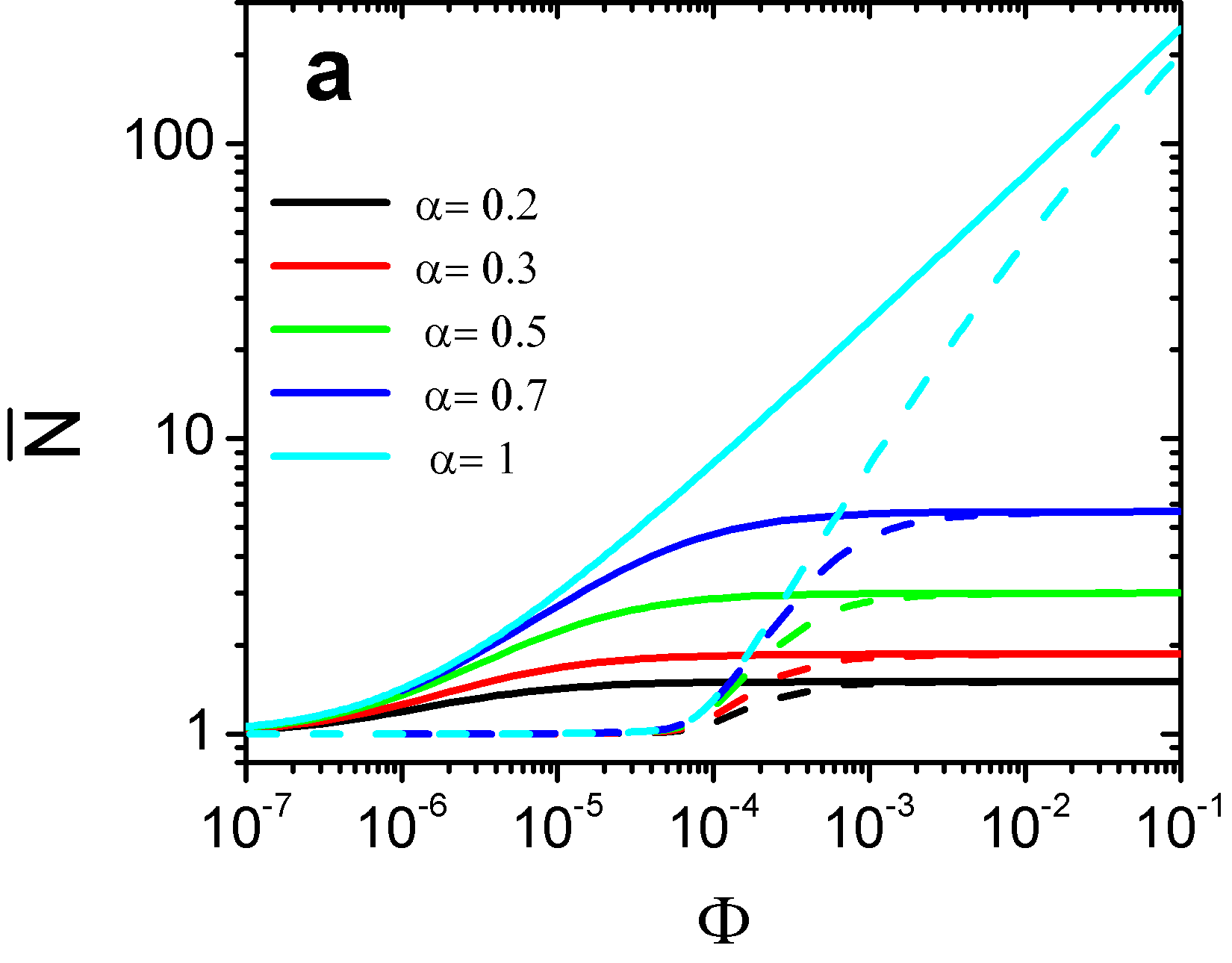}
  \includegraphics[width=.95\linewidth]{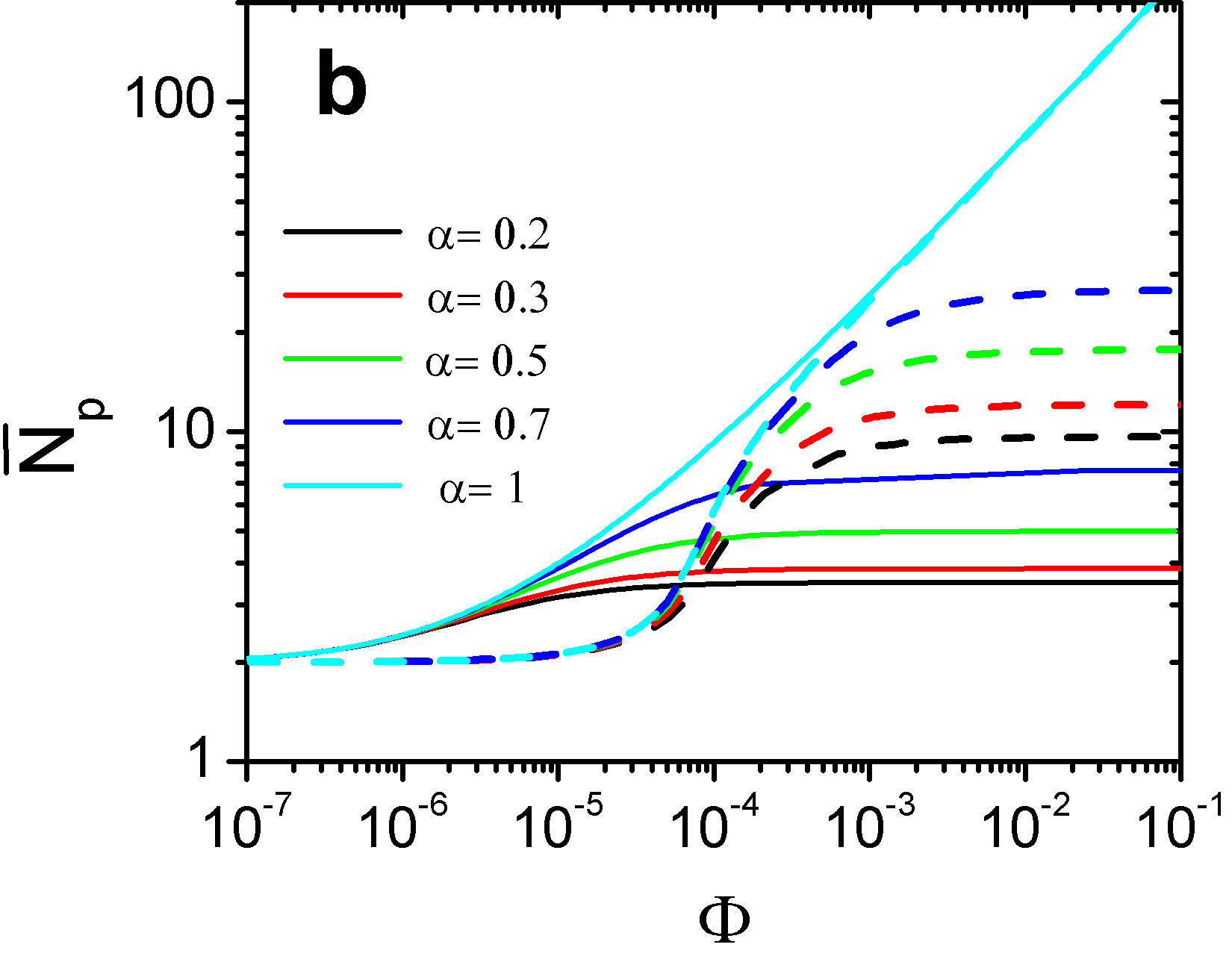}
     \caption{ The  mean degree of polymerization   defined by (a)  Eq. \eqref{eq:meandegree} and (b) Eq. \eqref{eq:meandegree1} as a function of  overall  concentration  of monomers, $\Phi$, at  different stoichiometric ratios $\alpha=\Phi_A/\Phi_B <1$, as given in the legends. The assumed values of the binding free energies are $b_{AA}=b_{BB} \rightarrow -\infty$ and $b_{AB}=b_{BA}=14$ yielding  $J \rightarrow - \infty$.  The solid  and dashed lines exhibit the results for  activation free energies $ a_A=a_B=a=0 $ and $4$, respectively.      
     }  
  \label{fig:Na_alt}
\end{figure}

Likewise, we also estimate the  maximum mean degree of polymerization $\overline{N}^{\text{max}}$ in the saturation regime for $J\rightarrow -\infty$.    
$\overline{N}^{\text{max}}$ is independent of the activation free energy and only depends on the stoichiometric ratio  
  \begin{equation}
\overline{N}^{\text{max}}=  \frac{1+\alpha}{1-\alpha },
\label{fig:Na_max}
  \end{equation}
  which is again only valid for $0<\alpha <1$. The results for $ \alpha > 1$ can be obtained by a transformation of the form $\alpha \rightarrow 1/\alpha$.
 Fig. \ref{fig:Na_alt}a presents the mean degree of polymerization $\overline{N}$ versus the overall volume fraction of monomers $\Phi$ at different  stoichiometric ratios $\alpha$ and for two different values of the activation free  energy. In perfect agreement with our theoretical estimate for  $\alpha<1$, the $\overline{N}$  at high concentrations saturates to a value that is given by 
  Eq. \eqref{fig:Na_max}. Only for the $\alpha=1$ case,  the mean degree of polymerization grows indefinitely with $\Phi$  and it is described by the well-known square-root law  \cite{Paulreview} at large concentrations.

  The mean degree of polymerization $\overline{N}$  defined by  Eq.~\eqref{eq:meandegree} is frequently used in  theoretical calculations \cite{PaulrevSMP,Paulreview} and it is of relevance to light scattering measurements. However, monomers are not considered in the average length of assemblies determined from transmission electron microscopy measurements. Therefore, it is  useful to calculate a mean polymerization degree where only assemblies with  $N\geq 2$ are included in the averaging. We denote such an average by $  \overline{N}_p$ defined as
  \begin{equation}
\label{eq:meandegree1}
  \overline{N}_p \equiv \frac{\sum_{N=2}^\infty N \rho(N)}{\vphantom{\sum^2}\sum_{N=2}^\infty \rho(N)}.
\end{equation}
 We have plotted $\overline{N}_p$ as a function of  $\Phi$ in Fig. \ref{fig:Na_alt}b. Similar to $\overline{N}$, it saturates to a maximum value $\overline{N}_p^{\text{max}}$ for $\alpha <1$ but its value now depends on the activation free energy.  It is larger for cooperative copolymerization $a>0$. The dependence of   $\overline{N}_p$ on stoichiometric ratio shows a good agreement with the  experimentally reported trend for the the average length of assemblies.~\cite{Ahlers2}  We can also find an analytical expression for it given as  
   \begin{equation}
\label{eq:Npmax}
  \overline{N}_p^{\text{max}}=\frac{2-\alpha +\sqrt{1-4 (\alpha -1) \alpha  \left(e^{\text{a}}-1\right)}}{1-\alpha},
 \end{equation}
 which is in perfect agreement with the numerical results. These results demonstrate  that  at a fixed overall volume fraction of monomers we can control the average length of polymers by  tuning the stoichiometric ratio of the two components.

\begin{figure}[t]
\centering
  \includegraphics[width=.47\textwidth]{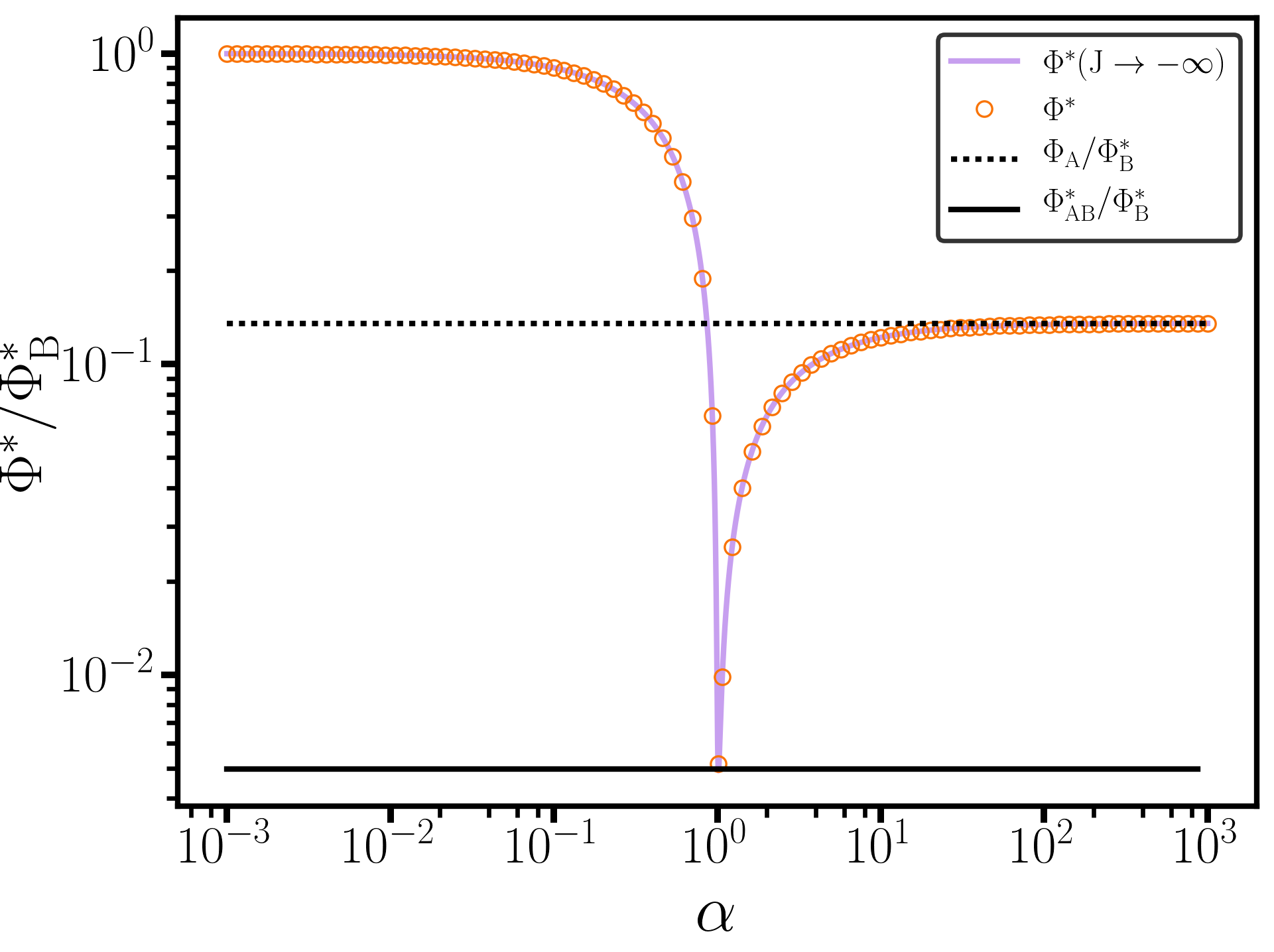}
  \caption{The critical concentration, $\Phi^*$, normalized by the critical concentration of species $B$, $\Phi^*_B$, as a function of the stoichiometric ratio $\alpha=\Phi_A/\Phi_B$.   The numerical solution of Eq.~\eqref{eq:critical} is displayed by orange open circles and the analytical result of Eq.~\eqref{eq:critconsimo} 
  is shown by a purple solid line.  The binding and activation free energies used are $b_{AA}=10$, $b_{BB}=8$, $b_{AB}=b_{BA}=14$,   and  ${a_A}=a_B=1.5$,  yielding  $\Phi^*_A=2.03 \times 10^{-4}$,  $\Phi^*_B=1.50 \times 10^{-3}$,  and  $\Phi^*_{AB}=7.45 \times 10^{-6}$. }
  \label{fig:critcon}
\end{figure}

\subsection {Critical concentration}
\label{sect_critC}

We obtain an analytical expression for  the critical concentration $\Phi^*$ in  the limit of $J \rightarrow -\infty$ for the case that $b_{ii} $ is positive and finite and the interspecies binding free energy is very large, {\it i.e.}, $b_{AB} \gg b_{ii} $. 
The critical concentration describes the transition from conditions of minimal assembly to those characterized by a strong polymerization.~\cite{Paulreview,nyrkova2000fibril}  At $\Phi^*$, the volume fraction of free monomers as a function of the overall concentration saturates, in other words  $\Lambda_+ \to 1$.  In coarse-grained models, $\Phi^*$  generally  depends on the  the effective binding and activation  free energies and the stoichiometric ratio. ~\cite{Paulreview,Sara}   As mentioned earlier, for monodisperse systems, the critical volume fraction reads $\Phi^*_i = \exp (-b_{ii} + a_i)$.  Notably, in the strong polymerization regime, {\it i.e.}, at high volume fractions $\Phi \gg  \Phi^*$,   the fraction of polymerized material $f$ obeys the simple relation $f=1-\Phi^*/\Phi$.~\cite{PaulrevSMP}

The conditions of polymerization transition for a bidisperse system are given by the following set of equations.~\cite{Sara}
\begin{eqnarray}
\label{eq:critical}
\Lambda_+(\mu_A^* , \mu_B^*)& \to 1,  \nonumber \\
\\  \nonumber 
\label{eq:critical1}
\left. \frac{\partial \Lambda_+}{\partial H}\right|_{\substack{\mu_A=\mu_A^*\\ \mu_B=\mu_B^*}} &= &\frac{\alpha -1 }{\alpha + 1},
\end{eqnarray}
where $\mu_A^*$ and $\mu_B^*$ are the values of the chemical potentials at the critical concentration.
Hence, the $\Phi^*$  is given by the total volume fraction of free monomers: 
\begin{equation}
\Phi^*(\alpha) =  \mathbf{e}^{\mu_A^*+a_A} +  \mathbf{e}^{\mu_B^*+a_B}.
\end{equation}
Note that this definition of critical concentration is only valid when $\Lambda_+ \to 1$ in the saturation limit of copolymerization. Therefore, it does not apply to the case of $b_{ii} \to -\infty$  and $\alpha \neq1$ where   $\Lambda_+$ at large concentrations saturates to $\Lambda_+^{\text{max}} <1$.

Taking into account the first two leading order terms of eigenvalues given in Eqs.~\eqref{eqEVtaylor}, we can find an analytical solution  for  Eqs.~\eqref{eq:critical} in the strongly negative coupling limit which results in 
\begin{equation}
 \Phi^*(\alpha)=\frac{(\Phi_A^*-\Phi_B^*) \left(\frac{\alpha - 1}{\alpha + 1}\right)+(\Phi_A^*+\Phi_B^*)\sqrt{\mathbf{e}^{4 J}+\left(\frac{\alpha - 1}{\alpha + 1}\right)^2 }}{1+\sqrt{\mathbf{e}^{4 J}+\left(\frac{\alpha - 1}{\alpha + 1}\right)^2 }}.
\label{eq:critconsimo}
\end{equation}
 For equal volume fractions of the two species, $\alpha=1$, Eq.~\eqref{eq:critconsimo} gives $\Phi^*(1)=(\Phi^*_A+\Phi^*_B)(1+e^{-2J})^{-1}$.  Especially, for the case of equal activation free energies, $a_A=a_B=a$, it can be simplified to $\Phi^*(1) \approx 2 \e{-b_{AB}+a} $  which is identical to $\Phi^*_{AB}$ defined earlier for  strictly alternating copolymers.  For small and large  values of $\alpha$, the critical concentration simplifies to
\begin{equation}
\Phi^*=
\begin{cases}
\Phi_B^*(1-\alpha) &\quad \alpha \ll 1  ,
\\
\Phi_A^*(1-\alpha^{-1}) &\quad\alpha \gg 1,
\end{cases}
\label{eq:critsimple}
\end{equation}
where $\Phi_A^*$ and $\Phi_B^*$ are the critical concentrations of homopolymers composed of $A$ and $B$ species, respectively. 
These equations show that  $\Phi^*$ approaches the critical concentration of the homopolymer of the more abundant species, in the appropriate limit. 

 To investigate the validity of Eq.~\eqref{eq:critconsimo}, we numerically solve the  set of Eqs.~\eqref{eq:critical} for a sufficiently negative coupling constant  $J=-5/2$ to obtain  $\Phi^*(\alpha)$ as plotted in Fig.~\ref{fig:critcon}.   For all  the stoichiometric values $\alpha$, we find  excellent  agreement between  the numerical and analytical solutions.  Additionally, we find that the strong coupling limit critical concentration, described by Eq.~\eqref{eq:critconsimo}, provides a good description of numerical results for all  coupling constants  $J<-1$.

\section{ Copolymerization behavior predominated by alternating configurations } 
\label{sec:results}

Having discussed the copolymerization with strictly alternating order, {\it i.e.}, antiferromagnetic regime in the $ J \rightarrow -\infty$ limit, we  obtain the numerical solution of  mass-balance equations   for a finite and negative coupling constant. We investigate the polymerization behavior and composition of
copolymers as a function of  overall volume fraction and stoichiometric ratio. Especially, we introduce an order parameter which quantifies the overall fraction of polymers with  perfect alternating order.


\subsection{Fraction of polymerized material and mean degree of polymerization}
\label{sec:fracMpolform}

\begin{figure}[htp]
\centering
    \includegraphics[width=.5\textwidth]{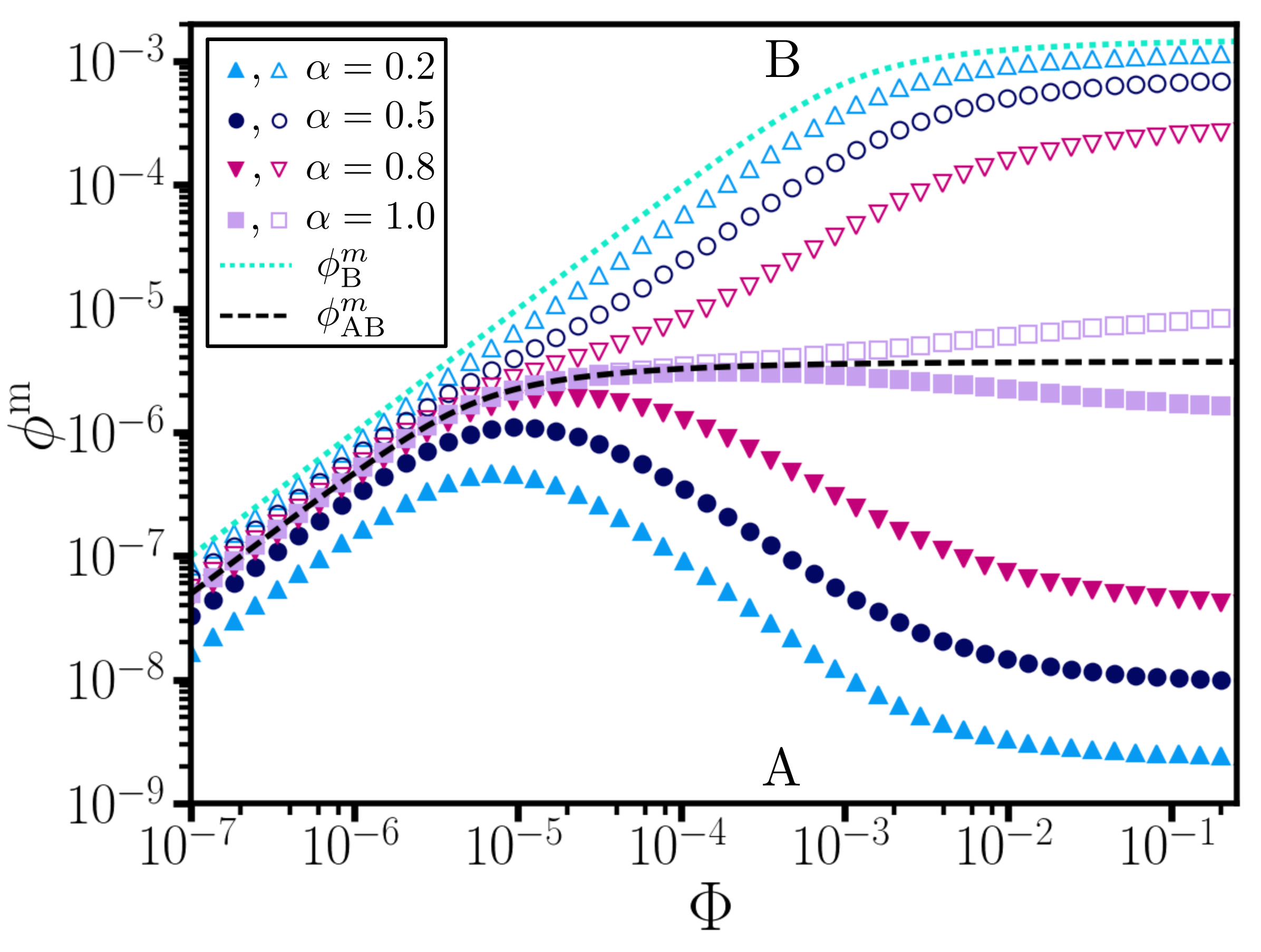}
  \caption{The volume fraction of free monomers of  $A$ species   (closed
symbols)  and  $B$ species  (open symbols)   as a function of  the overall volume fraction  $\Phi=\Phi_A+\Phi_B$ at different stoichiometric ratios  $\alpha=\Phi_A/\Phi_B $, as given in the legend. The  dotted and dashed lines show  the free monomer volume fractions for the special cases of  B-homopolymers $\phi^m_B$, and strictly alternating $AB$-copolymers $\phi^m_{AB}$, respectively. The assumed values of the activation  and  binding  free energies   are  ${a_A}={a_B}=1.5$ and $b_{AA}=10$, $b_{BB}=8$ and $b_{AB}=b_{BA}=14$  resulting in a negative coupling  constant $J=-5/2$. Note that the case $\alpha = 1$ loses A-B symmetry because the $b_{AA} > b_{BB}$.}
  \label{fig:freemono}
\end{figure}

\begin{figure}[t]
\centering
  \includegraphics[width=.45\textwidth]{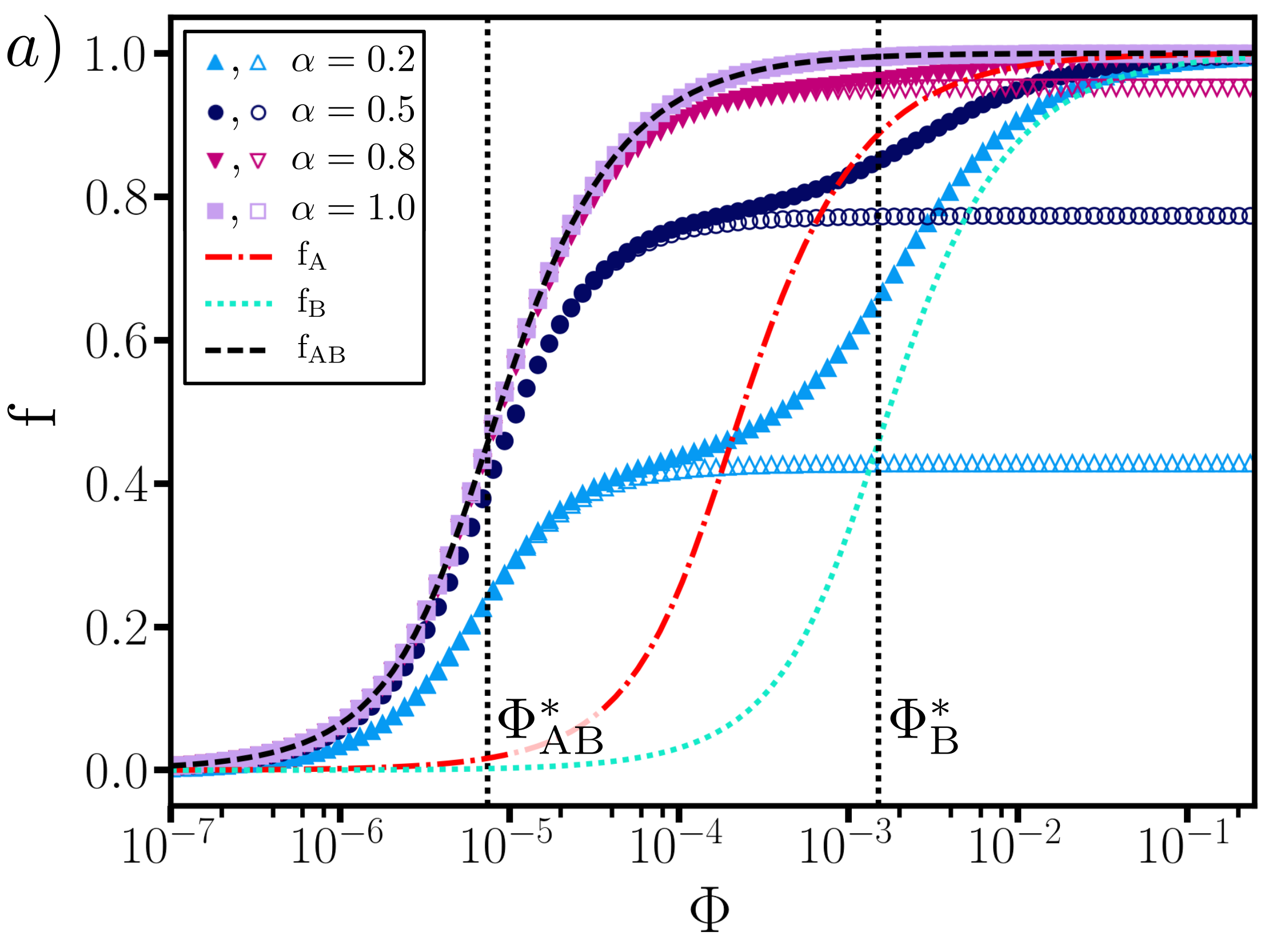}
   \includegraphics[width=.45\textwidth]{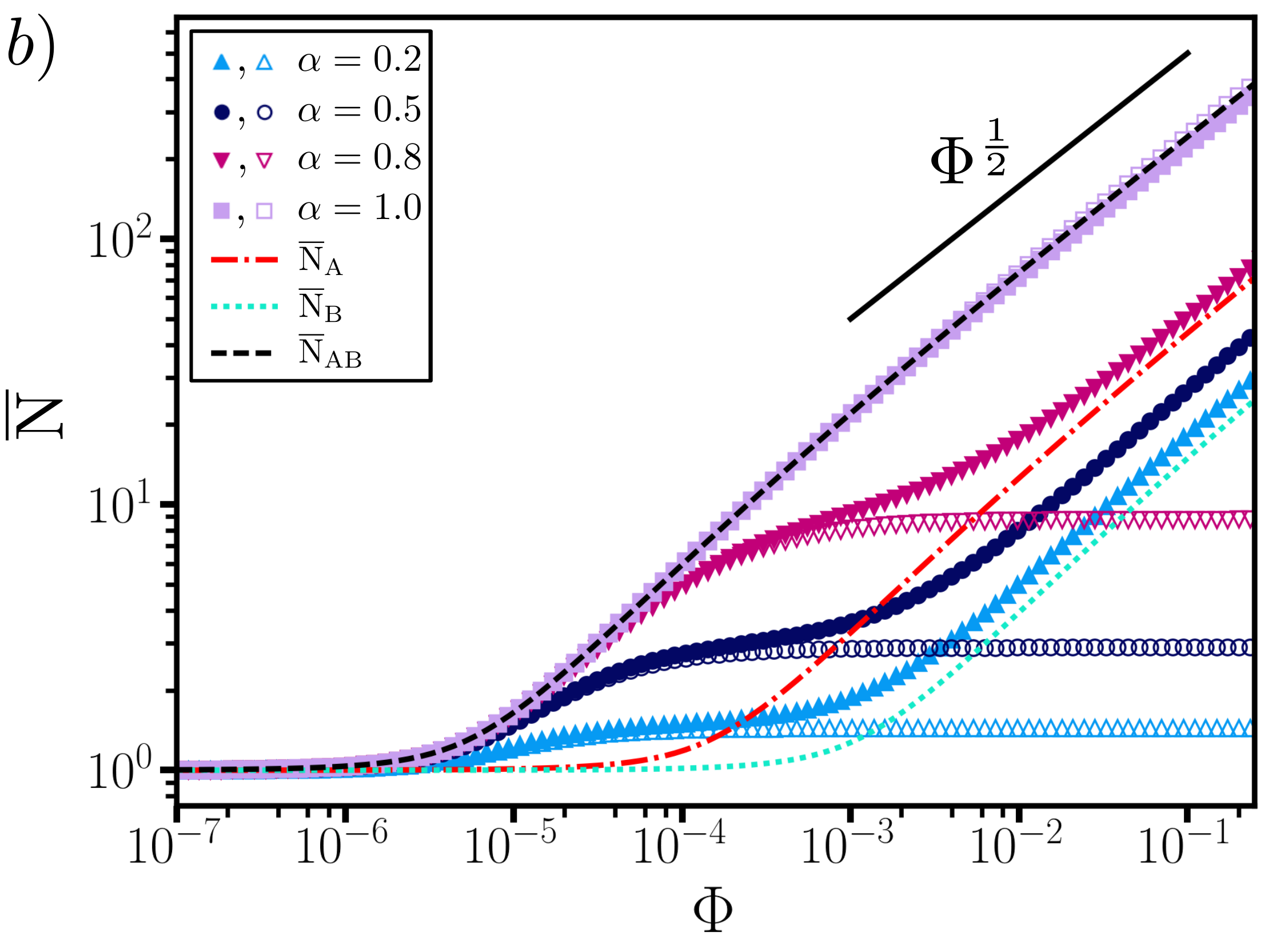}
     \caption{a) Fraction of polymerized material $f$ and b) Mean degree of polymerization $\overline{N}$  as a function of the overall volume fraction of dissolved monomers $\Phi=\Phi_A+\Phi_B$ at different stoichiometric ratios as given in the legends. The
     filled symbols correspond to polymerization curves of  two component self-assemblies with activation  free energies ${a_A}={a_B}=1.5$ and 
    binding free energies $b_{AA}=10$, $b_{BB}=8$ and $b_{AB}=b_{BA}=14$, yielding 
   a coupling constant $J=-5/2$. The results for homopolymers consisting of  $A$ species ($f_A$ and $\overline{N}_A$, dash-dotted lines) and of  $B$ species ($f_B$ and $\overline{N}_B$, dotted lines) and the strictly alternating copolymer consisting of repeat units of $(AB)$ ($f_{AB}$ and $\overline{N}_{AB}$, dashed lines) are also presented.  Furthermore, the  open symbols show $f$ and $\overline{N}$ 
   in the strongly negative coupling limit, $J  \to -\infty$ when $b_{ii} \to -\infty$, computed from the numerical solutions of Eq. (\ref{eq:totvolfraclim}) and (\ref{eq:diffvolfraclim}), with otherwise identical parameters.}
    \label{fig:frac}
   \end{figure}

 We  fix the values of the free energies such that  $b_{AB} > b_{AA} > b_{BB}$ and we focus on the polymerization behavior for stoichiometric ratios $ 0 < \alpha  \le 1$. We note that the labels of particles, $A$ and $B$, are a priori arbitrary and the above conditions determine the specific labels of the two species. Therefore, the results for $\alpha > 1$  can be simply obtained through an interchange of  species labels and our  presented analysis is without loss of generality.    We  set  the binding free energy values to $b_{AA}=10$, $b_{BB}=8$ and $b_{AB}=b_{BA}=14$. These values  give rise to $J=-5/2$  which is sufficiently negative to expect the alternating order to be the predominant morphology. Moreover, we choose equal activation free energy values for the two species  $a_A = a_B = 1.5$ to reduce the number of parameters.  They are chosen  small enough to allow us to obtain accurate numerical results.  As discussed in section \ref{sec:strict_alt}, the activation free energy value affects the sharpness of the transition from the monomeric to the polymeric dominated regime, but does not alter  the overall self-assembly behavior.

 We  determine the chemical potentials $\mu_i$ by solving the mass-balance equations Eqs.~\eqref{eq:Totalvolfrac} and  ~\eqref{eq:Diffvolfrac} numerically from which we calculate the  volume fraction of free monomers in  the solution as $\phi_i^{m}= \exp(\mu_i+a_i)$.
In Fig.~\ref{fig:freemono}, we have depicted the $\phi_i^{m}$ as a function of  the overall volume fraction $\Phi$ at different  stoichiometric ratios $ 0<\alpha \le 1$. We find that at very low concentrations, $\Phi <10^{-5}$, the ratio of free monomer volume fractions $\alpha^f \equiv \phi_A^{m}/\phi_B^{m} $ is almost identical to $\alpha$. However, at larger concentrations and for $\alpha<1$, the value of $\phi_A^{m}$  drops considerably such that  $\alpha^f  \ll \alpha$. Even for the case of $\alpha=1$, we notice that  $\Phi^m_A  < \Phi^m_B$  when $\Phi > 10^{-4}$ because of a greater tendency of $A$ species to homopolymerize.
   

We next examine the fraction of polymerized material $f$ as a function of the overall volume fraction $\Phi$   presented in Fig.~\ref{fig:frac}a for various stoichiometric ratios $0 < \alpha   \le   1$. Here,  $f_A$ and $f_B$  represent the fraction of polymerized material for homopolymers consisting of $A$ and $B$
species, respectively. They correspond to the results for the limiting cases of   $\alpha=\infty$  and  $\alpha=0$   that are determined from  the mass balance equation given in  Eq.~\eqref{eq:mono}.  For comparison, we  have also presented
the fraction of polymerized material  in the strongly negative coupling limit, {\it i.e.}, $b_{BB}=b_{AA} \to -\infty$ and  with otherwise  identical free energy parameters. These curves, depicted by open symbols,  are obtained by solving the mass-balance Eqs.~\eqref{eq:totvolfraclim} and \eqref{eq:diffvolfraclim}.  $f_{AB}$, shown by the dashed line, depicts the results for the special case of $\alpha=1$  (quasi-homopolymers made of  $(AB)$ monomers)  and it  is  obtained from solving   Eq.~\eqref{eq:AB-polymer}. Figure \ref{fig:frac}a shows that the fraction of polymerized material  is bounded  by the polymerization 
curves  $f_{AB}$ and  $f_{B} (f_A)$ for $\alpha  <  1$ ($\alpha >  1$). The polymerization behavior  of copolymers with a finite negative coupling constant at high concentrations is  notably different from that of strictly alternating copolymers. In the latter case, $f$ saturates to a value $f^{\text{max}} (\alpha) <1$, unless $\alpha =1$.
     However, at lower concentrations the polymerization curves are identical to those of  strictly alternating copolymers. This agreement suggests that copolymers at low volume fractions have a predominantly alternating order.


Our conclusions are further supported by the  concentration dependence of mean degree of polymerization presented in Fig. \ref{fig:frac}b. We note that at low and intermediate concentrations, the  $\overline{N}$ of copolymers with finite coupling constant (filled symbols) agrees with those in the strongly negative coupling limit (open symbols). This figure further confirms  that at  the early stages of polymerization short alternating polymers are formed.
At very low volume fractions $\Phi \lesssim \Phi^*_{AB}$, the solution is mainly in a monomer-dominated state and $\overline{N}\approx 1$. At the intermediate volume fractions, $\overline{N}$ monotonically increases with a slope that depends on  $\alpha$ and $\Phi$. 
At high $\Phi$ the polymer assemblies follow the well-known square root law found for homopolymers~\cite{Paulreview} but with a coefficient that increases with $\alpha$. 

\subsection{Composition of supramolecular copolymers}
%

%
\begin{figure*}[ht]
\centering
  \includegraphics[width=.45\textwidth]{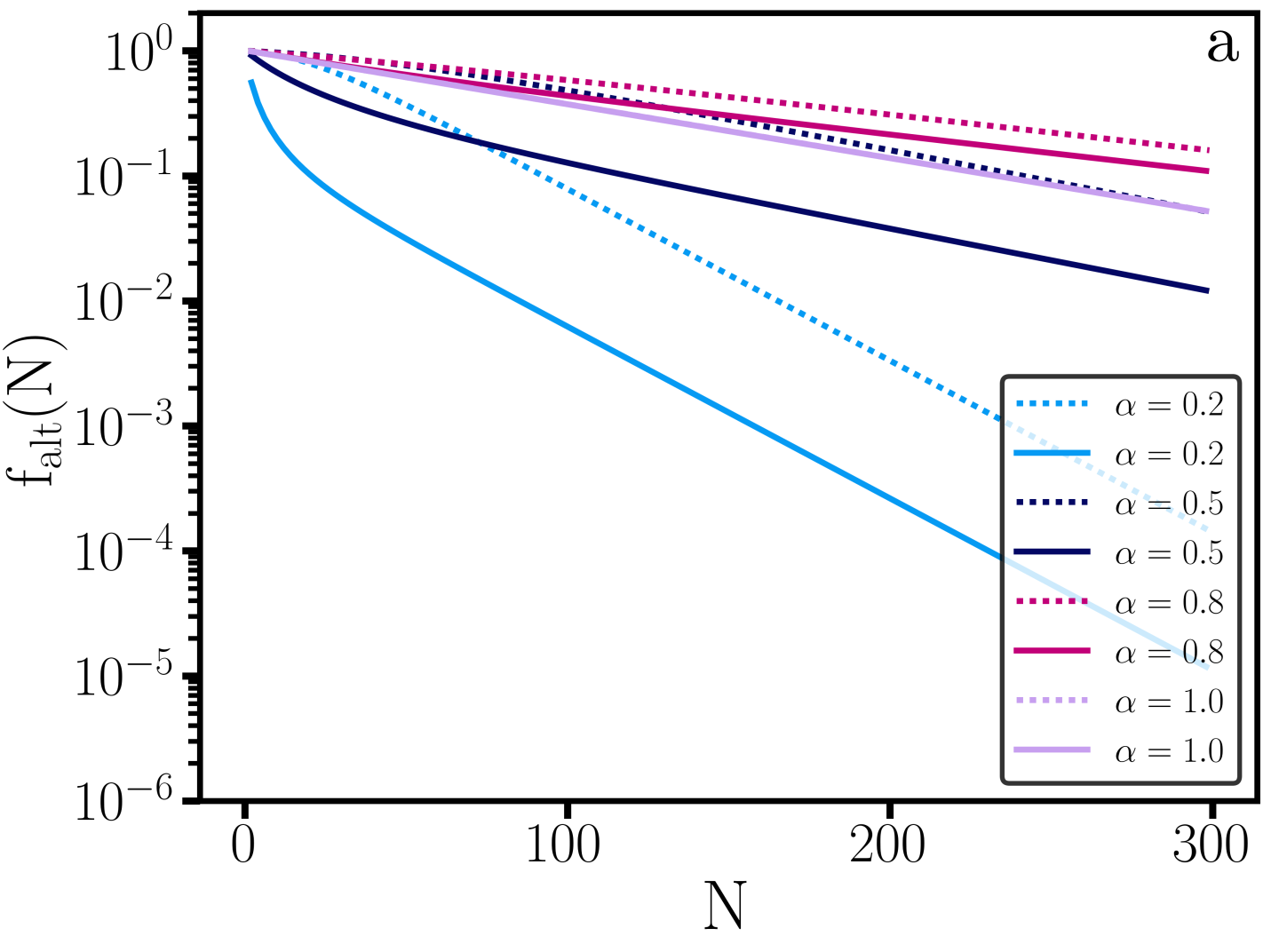}
  \includegraphics[width=.45\textwidth]{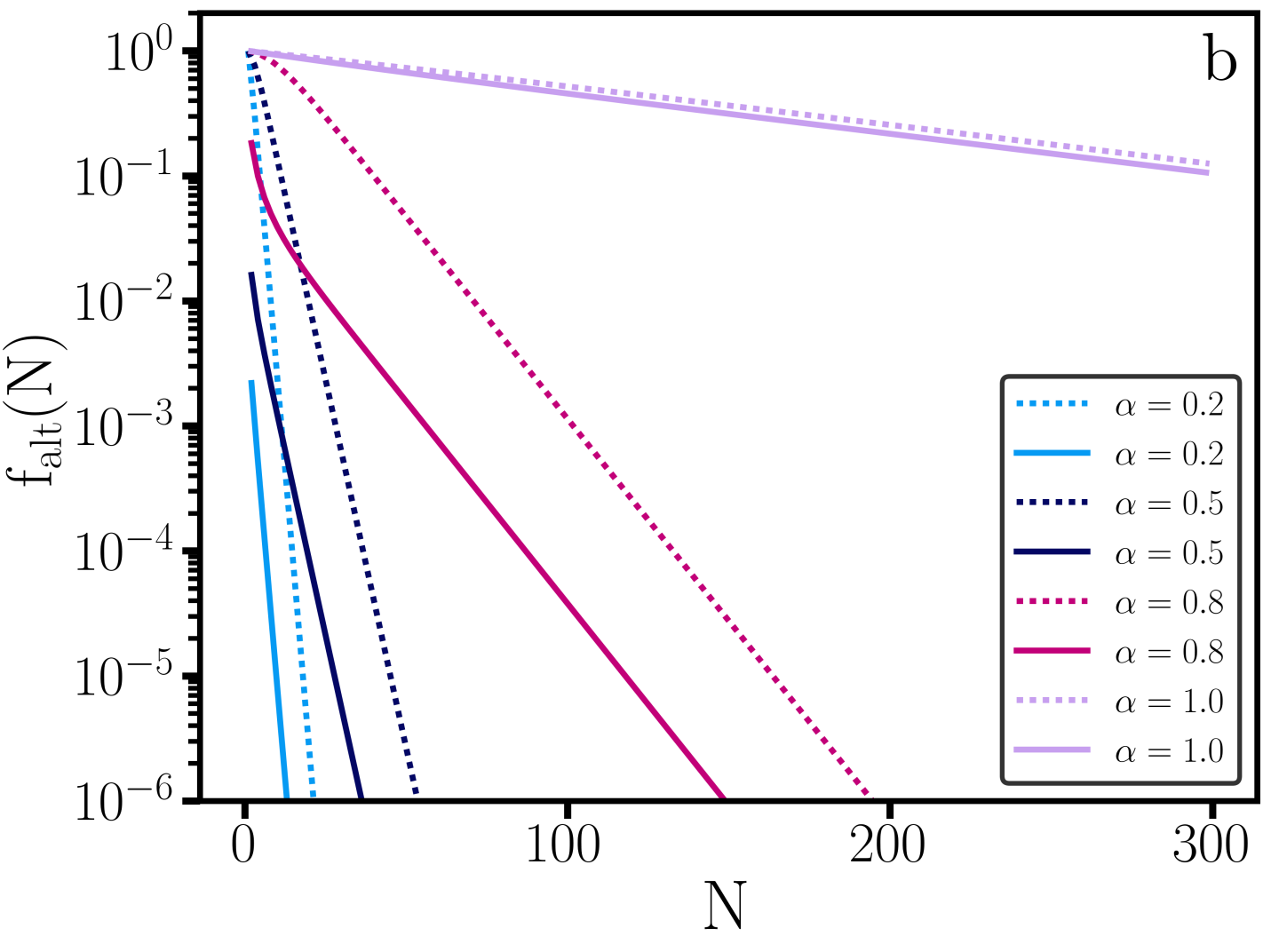}
  \includegraphics[width=.45\textwidth]{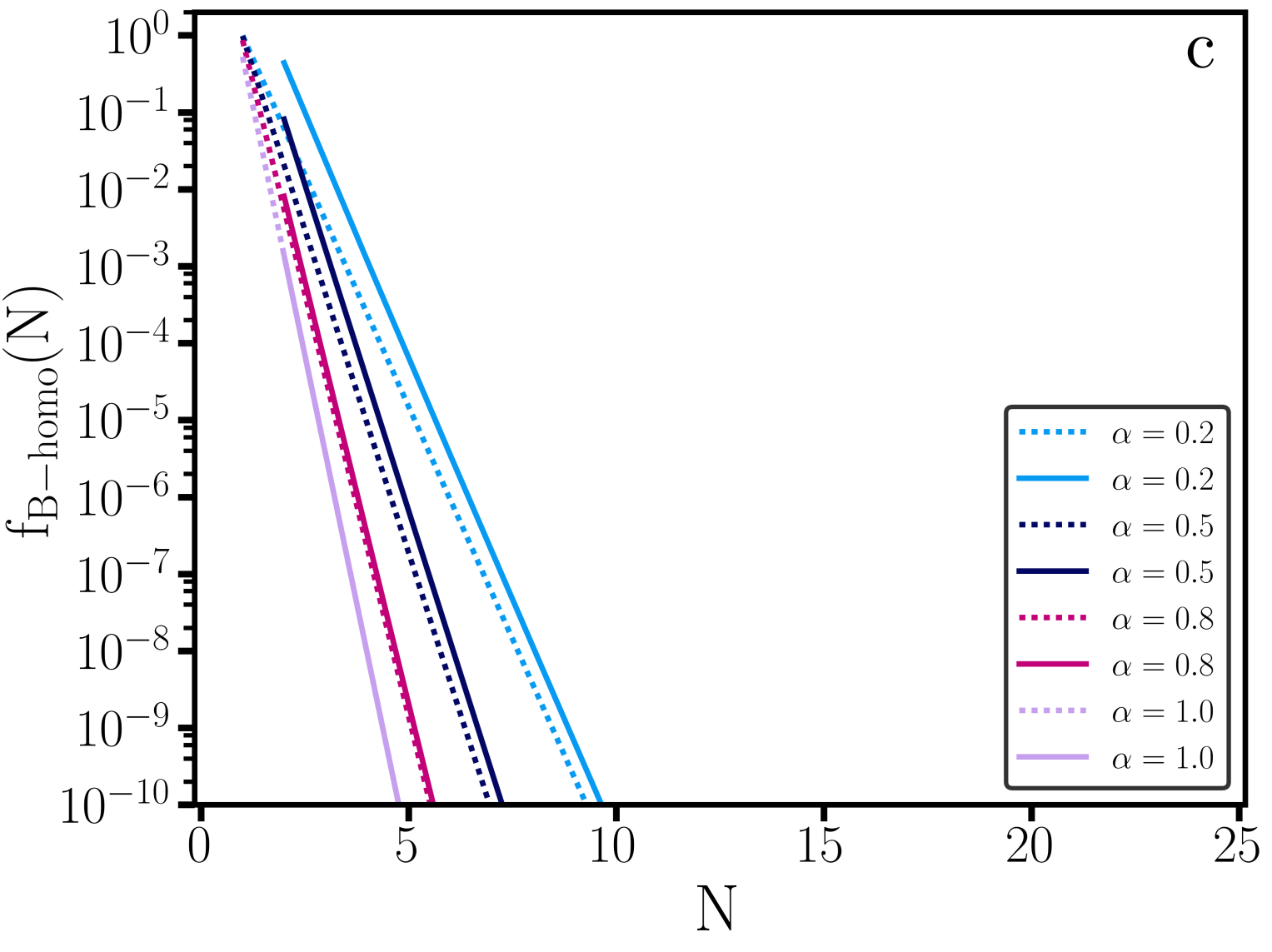}
  \includegraphics[width=.45\textwidth]{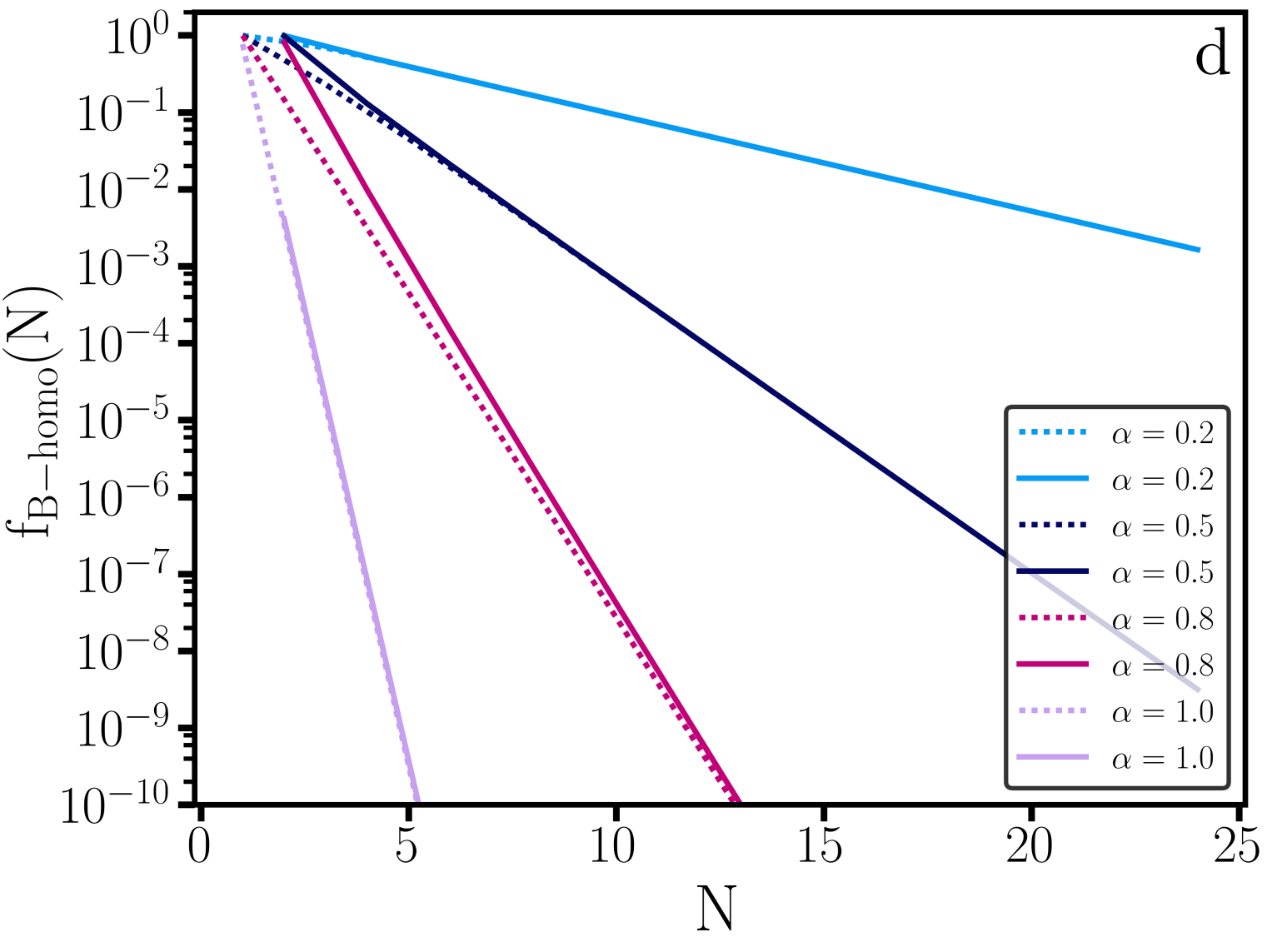}
  \caption{(a) and (b) the  fraction of copolymers of length $N$ with perfect alternating order relative to all  polymers with the same length, $f_{\text{alt}}$, as defined by Eq. \eqref{eq:f_altN},  (c) and (d) the
  fraction of  $N$-mer $B$-homopolymers  relative to all copolymers with the same length $f_{\text{B-homo}}(N)$, as defined by Eq. \eqref{eq:f_Bhomo}, plotted against $N$  at different stoichiometric ratios $\alpha$ given in the legends. The dotted  and solid  lines correspond to polymers with odd and even degrees of polymerization, respectively. The overall monomer concentration is
   $\Phi=10^{-4}$ in panels (a) and (c) is  and $\Phi=10^{-2}$  in panels (b) and (d). The values of the activation and the binding free energies are ${a_A}={a_B}=1.5$ and $b_{AA}=10$, $b_{BB}=8$ and $b_{AB}=b_{BA}=14$,  resulting in  $J=-5/2$. }
  \label{fig:f_althomoN}
\end{figure*}
In this subsection we focus on the composition of copolymers. To provide a quantitative insight, we  determine the fraction of  perfectly alternating copolymers and $B$-homopolymers as a function of concentration and assembly length. 
For a given supramolecular polymer length $N$, the fraction of strictly alternating copolymers
is determined by 
\begin{equation} 
\label{eq:f_altN}
 f_{\text{alt}}(N)=\frac{\sum_{\text{config}}\exp[-\mathcal{H}_{alt}(N)]}{Z_{N}},
\end{equation}
  where $Z_N$ is the copolymers partition function given by Eq.~\eqref{eqpartition} and $\mathcal{H}_{alt}(N)$ is 
  the  free energy of an alternating configuration of size $N$. Summing over all alternating configurations in the numerator  yields $Z_{N>1}^\text{even}$ and $Z_{N>1}^\text{odd}$, defined by Eqs.~\eqref{eqZeven} and \eqref{eqZodd}, for even and odd degrees of polymerization, respectively.  From now on, we refer to both  as  $Z_N^{alt}$.  
Likewise,  the fraction of  $B$-homopolymers  is obtained as 
 \begin{equation}
 \label{eq:f_BhomoN}
  f_{\text{B-homo}}(N)= \frac{\exp[-\mathcal{H}_{\text{B-homo}}]}{Z_{N}} =\frac{  \exp[ \mu_B N+(N-1)b_{BB}] }{Z_{N}},
\end{equation}
 where $\mathcal{H}_{\text{B-homo}}$ is the free energy of a homopolymer of size $N$.

 Fig.~\ref{fig:f_althomoN}a and   \ref{fig:f_althomoN}b present the fraction of strictly alternating copolymers $f_{\text{alt}}(N)$ as a function of the assembly length $N$ at two concentrations $\Phi=10^{-4}$ and $10^{-2}$ and for different stoichiometric ratios.  Two important conclusions can be drawn from these results. Firstly, the fraction of alternating polymers with even degree of polymerization is smaller than fraction of those with odd degree of polymerization, except for the case of $\alpha=1$ and $\Phi=10^{-4}$. Notably, the difference between populations of odd and even numbered alternating copolymers is larger for smaller $\alpha$  and it reflects the lack of the scarcer $A$ species. The more abundant $B$ species can have a greater contribution to polymerization by  forming alternating copolymers of the form $B(AB)_{2N}$. Secondly, the fraction of perfectly alternating copolymers decreases with $N$ and this decrease is stronger for smaller  stoichiometric ratios. Interestingly, the  $f_{\text{alt}}(N)$ for stoichiometric ratios  $\alpha=0.2$ and 0.5  exhibits a two-step decay. An initial rapid decay for short copolymers and a second slower exponential decrease for longer assemblies. At the higher concentration, where the fraction of polymerized material is larger, the fraction of alternating copolymers shows a stronger decrease with $N$, even for the case of  equal concentrations, $\alpha=1$. 
 
 The  observed behavior is consistent with the picture  arising from the Ising model that its correlation length  \cite{goldenfeld1992} can be estimated as $\xi_0 \approx \exp(-2J)/2 \approx 74$. At low and intermediate concentrations where $\overline{N} < \xi_0$,  the short assemblies have a nearly  perfect alternating order. However, upon increase of concentration and growth of mean degree of polymerization, for  long assemblies with $N \gg \xi_0$  the combinatorial factor can benefit from the formation of alternating copolymers  with defects. Additionally, at sufficiently high $\Phi$ for $\alpha <1$ the excess of $B$ species can form homopolymers.
 The fraction of  $B$-homopolymers  versus $N$ is shown in Fig.~\ref{fig:f_althomoN}c and   \ref{fig:f_althomoN}d for $\Phi=10^{-4}$ and $\Phi=10^{-2}$, respectively.  As can be seen at the lower concentration, there exist almost no homopolymers whereas at the higher concentration for $\alpha <1$, a notable fraction of homopolymers appear. Evidently, at the smaller $\alpha $ with a larger excess of $B$ species, longer homopolymers are more probable.
  
Next, we calculate the  total fraction  of perfectly alternating copolymers and homopolymers as a function of  the overall concentration of monomers.  The  total fraction of perfectly alternating copolymers can be  obtained as
\begin{eqnarray}
\label{eq:f_alt}
f_ {\text{alt}} ^{\text{tot}} \equiv \frac{\sum_{N=2}^\infty   \rho(N) f_{\text{alt}}(N)}{\vphantom{\sum^2}\sum_{N=2}^\infty \rho(N)}
= \frac{\sum_{N=2}^\infty  Z_N^{alt}}{\vphantom{\sum^2}\sum_{N=2}^\infty Z_N}.
\end{eqnarray}
  Using formulas for convergent geometric series, both sums can be evaluated to express $f_ {\text{alt}} ^{\text{tot}}$ in terms of the chemical potentials
$\mu_i$.  Similarly, the   total fraction of  $B$-homopolymers is given by 
\begin{eqnarray}
\label{eq:f_Bhomo}
f_ {\text{B-homo}} ^{\text{tot}} &\equiv& \frac{\sum_{N=2}^\infty   \rho(N) f_{\text{B-homo}}(N)}{\vphantom{\sum^2}\sum_{N=2}^\infty \rho(N)} \\ \nonumber
&=& \frac{\sum_{N=2}^\infty   \exp[ \mu_B N+(N-1)b_{BB}]}{\vphantom{\sum^2}\sum_{N=2}^\infty Z_N}, 
\end{eqnarray}
that can be straightforwardly evaluated.

\begin{figure}[ht]
\centering
  \includegraphics[width=.5\textwidth]{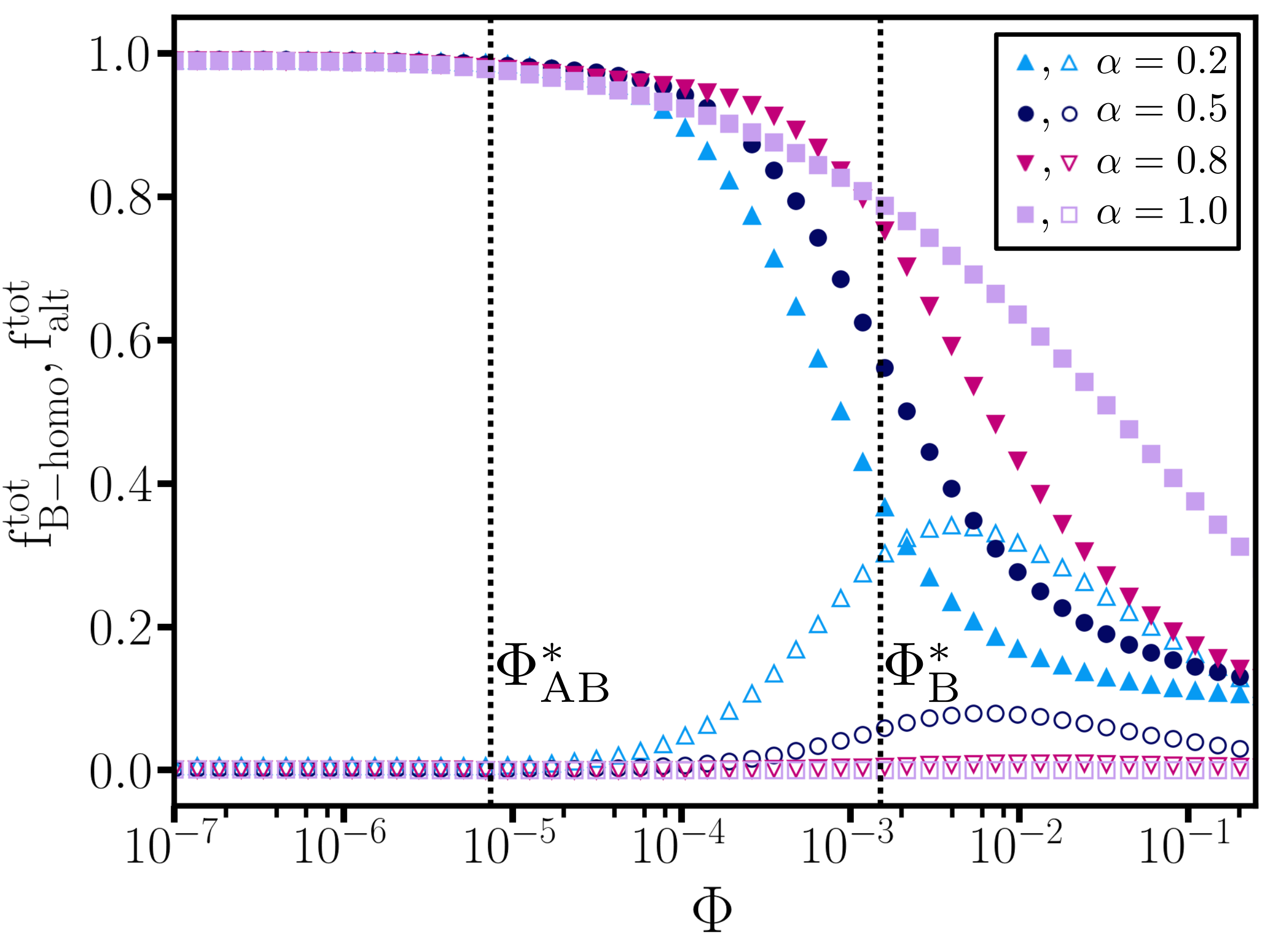}
  \caption{The total fraction of copolymers with 
   perfect alternating  order $f_ {\text{alt}} ^{\text{tot}}$ (closed symbols) and the  total fraction of homopolymers  
   consisting of  $B$  species $f_ {\text{B-homo}} ^{\text{tot}}$  (open symbols) as a function of the overall volume fraction of dissolved monomers $\Phi=\Phi_A+\Phi_B$, at    different stoichiometric ratios $\alpha=\Phi_A/\Phi_B$  as given in the legend .
   The values of the activation and binding free energies are ${a_A}={a_B}=1.5$ and $b_{AA}=10$, $b_{BB}=8$ and $b_{AB}=b_{BA}=14$,  resulting in     $J=-5/2$. }
 
  \label{fig:f_copolymer}
\end{figure}

Fig.~\ref{fig:f_copolymer}  shows $f_ {\text{alt}} ^{\text{tot}}$ and $f_ {\text{B-homo}} ^{\text{tot}}$ as a function of the overall  volume fraction  at different stoichiometric ratios. At low volume fractions $\Phi  \lesssim \Phi^*_{AB}$, where  most of the material is in monomeric form and only few short polymers are formed,  $f_ {\text{alt}} ^{\text{tot}}\approx 1$ and $f_ {\text{B-homo}} ^{\text{tot}} \approx 0$. Hence, we conclude that at low $\Phi$ the majority of  assemblies are in an alternating configuration.
 For $\Phi  >  \Phi^*_{AB}$, the fraction of perfectly alternating copolymers  monotonically decreases  with the concentration whereas 
the fraction of B-homopolymers versus concentration exhibits a maximum for $\alpha <1$. The maximum in $f_ {\text{B-homo}} ^{\text{tot}} (\Phi)$
appears at  volume fractions where the concentration of excess $B$-monomers becomes comparable to $\Phi_B^*$. 
At high  $\Phi$, $f_ {\text{alt}} ^{\text{tot}}+  f_ {\text{B-homo}} ^{\text{tot}} <1$ which 
 shows that the system contains alternating copolymer and homopolymer configurations with defects in addition to strictly alternating copolymers and pure homopolymers. The existence of such configurations reflects that the mixing entropy overcomes the energetically unfavorable compositions, particularly for long assemblies where the combinatorial factor is larger. 

\section{Link to experimental observations}
\label{sec:exp}

In this section we rationalize the experimental observations on the copolymerization of positively and negatively
charged comonomers ~\cite{Frisch,Ahlers,Ahlers1} on the basis of our theoretical insights.
  As briefly discussed in section \ref{sec:intro}, the comonomer species with complementary charges contain three identical amphiphilic oligopeptide arms that possess a $C_3$ symmetry. Each arm has 2 charged groups and therefore a monomer has at most a total charge of $6$ $e$, depending on the pH conditions. The oligopeptide design of each arm is based on a phenyl alanine and methionine sequence for the hydrophobic amino acid, alternated with a cationic
lysine (blue comonomer A in Fig.~\ref{fig1}) or anionic glutamic acid moiety (green comonomer B in Fig.~\ref{fig1}). 
The terminal hydrophilic dendritic triethylene glycol chains 
are introduced to guarantee a high colloidal stability of the copolymers in  a neutral buffer of physiological ionic strength. Further details on the synthesis of the $C_3$ symmetrical comonomers can be found in reference.~\cite{Engel}

 At neutral pH values, the two species are  oppositely charged and they copolymerize into linear aggregates.~\cite{Besenius}   
 The electrostatic interactions between the monomers  enhance the  binding between the $A$ and $B$ species with complementary charges and reduce the
 binding free energy between identical species. Therefore, the copolymerization of this system corresponds to the regime
 where interspecies binding is favored over  homopolymerization, leading  to a  strongly negative
 coupling constant  $J$ in the language of our theory.  To make the link between the theory and experiments more quantitative, we provide a rough estimate of the effective coupling constant.

 Due to the similar chemical architecture of the two species, we  assume that the bare binding free energies  in the uncharged state  are equal, {\it i.e.}, $b_{ij}=b_0$. In the charged state,  the binding free energy values are modified due to electrostatic interactions.   The  magnitude of electrostatic  free energy  between any two neighboring monomers is given by $b_\text{el}$  because the two species have equal  charge magnitudes. As a result, the effective binding free energies modify to $b_{AA}=b_{BB}=(b_0-b_\text{el})$ and $b_{AB}=b_{BA}=(b_0+b_\text{el})$, leading to a coupling constant $J=-b_\text{el}$; see Eq.~\eqref{eq:J}. We estimate the electrostatic contribution based on the Debye-H{\"u}ckel theory~\cite{DH} for  pair interactions between point-like charges in an electrolyte 
 as $b_\text{el}= \lambda_B  {q^2} \exp(-\ell/\lambda_D)/ \ell$,  where $q$ represents an effective charge valency and $\ell$ is the distance between the monomers.  $\lambda_B \equiv e^2 \big/ (4 \pi \epsilon k_BT)$ is the Bjerrum length in which $e$ is the elementary charge and $\epsilon$ is the solvent permittivity. $\lambda_D$ represents the Debye screening length, which depends on the ionic strength of the 
 buffer  solution.~\cite{DH} 

\begin{figure}[t]
\centering
  \includegraphics[width=.5\textwidth]{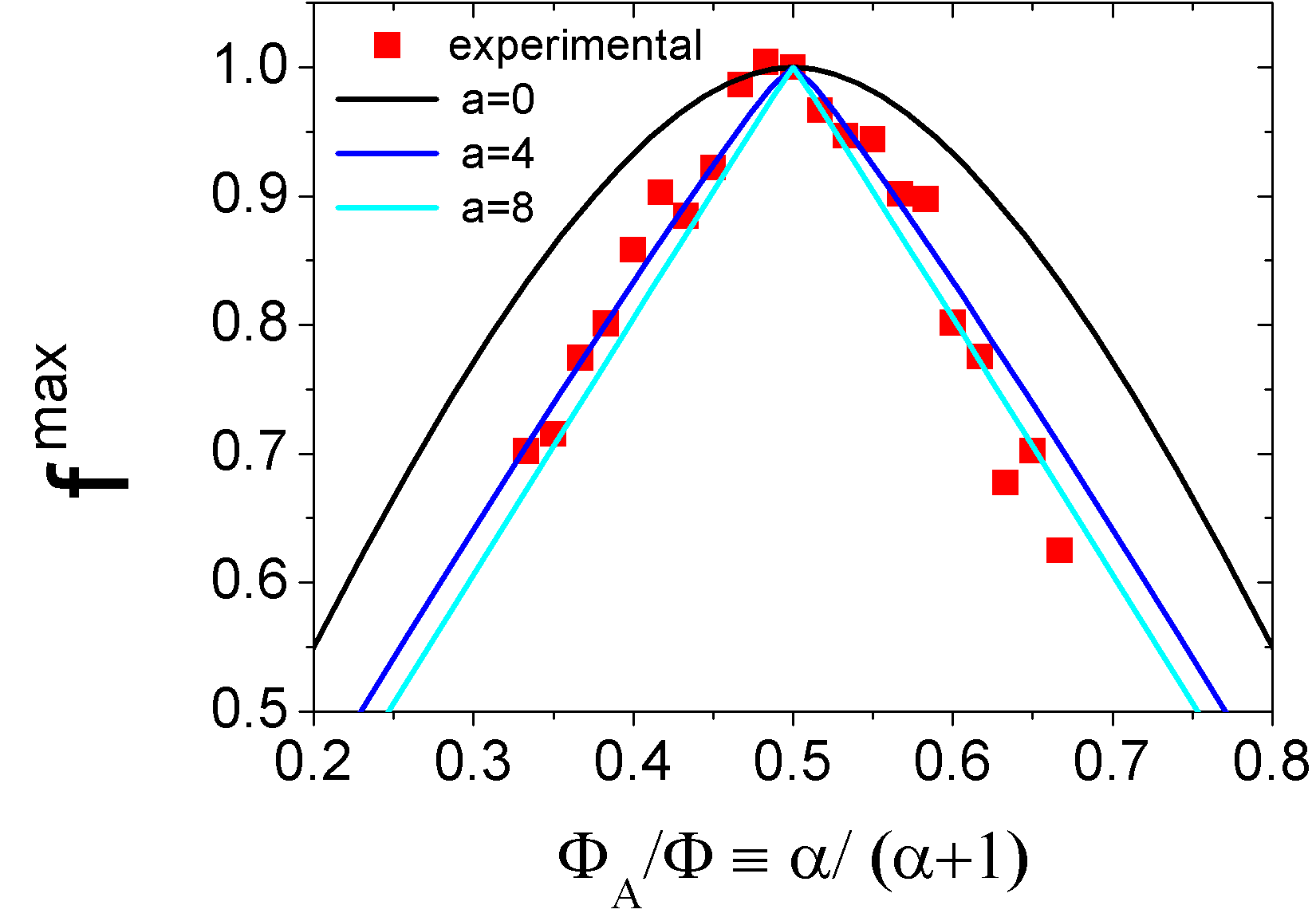}
  \caption{ Fraction of polymerized material in the saturation limit $f^{\text{max}}$ plotted against the  volume fraction of  $A$  species relative to the overall volume fraction of dissolved monomers, denoted by $\Phi_A/\Phi \equiv \alpha/(\alpha+1)$.   The experimental data are obtained from CD measurements at a wavelength of 220 nm for a mixture of glutamic acid derivatives (negatively charged) and  lysine derivatives (positively charged)
  as commoners  with a total concentration  of 60 $\mu$M at pH 6.0.  The  solid lines  show the theoretical prediction for $f^{\text{max}}$  given by Eq.~\eqref{eq:fmax}  at different activation free energies.}
   \label{fig:expfrac}
\end{figure}
%
For the system discussed above, we expect  $q < 6$ $e$, due to projection from the discotic surface charge density into a  point-like charge distribution along the linear assemblies and electrostatic screening resulting from counterion condensation.~\cite{PolPaul}  We estimate it to be in the range $ 3 < q <  6$ and the Debye screening length to be about $\lambda_D \approx 1$ nm under our experimental conditions.
 The Bjerrum length at room temperature is $\lambda_B \approx 0.7$ nm and the spacing between monomers is $\ell=0.47$ nm, on the basis of WAXS measurements.~\cite{Spitzer2017}   Using these values, we estimate the  electrostatic energy to be in the range  $ 8.3 < b_\text{el} < 33 $ leading to a large negative $J=-b_\text{el}$. Hence, the correlation length $3.8 \times 10^6 $ nm  $< \xi_0< 1.08 \times 10^{28}$ nm  is so large  that it favors strictly alternating copolymer configurations.  Accordingly, homopolymerization will  only  take place 
 if $b_0 > b_\text{el}$; otherwise, the electrostatic repulsion will disfavor any binding between identical species.
Therefore,  we expect that our experimental system to fall in the strongly  negative coupling regime; see Sec.~\ref{sec:Strongcoupling}.   
To test this hypothesis, we compare the maximum fraction of self-assembled material 
obtained from Eq. ~\eqref{eq:fmax} with the experimental values.



In the experiments, the fraction of polymerized material $f$ is obtained by circular dichroism (CD) spectroscopy where the normalized value for the molar circular dichroism $\Delta \epsilon$ is plotted as a function of the monomer stoichiometric ratio. We have previously shown that $\Delta \epsilon$ scales linearly with the fraction of polymerized material.~\cite{Frisch, Besenius,Ahlers} 
The feed titrations were performed in a 2 mm cuvette using separately prepared solutions of both monomers at pH 6.0 at a total monomer concentration of $6 \times 10^{-5}$ M, while monitoring the maximum of the negative CD band at a wavelength of 220 nm.
The fraction of polymerized material was estimated by normalizing the intensity of the polymer specific CD band at 220 nm, where 0 refers to the monomeric state and 1 to the fully polymerized state.

 We measured the value of the CD signals as a function of the relative volume fraction of  $A$ species with respect to  the overall volume fraction of monomers in the solution,  $\Phi_A/\Phi \equiv \alpha/ (\alpha+1)$.   
The experimentally estimated fraction of polymerized material in the saturation regime is shown  in Fig.~ \ref{fig:expfrac}. It  clearly shows that the fraction of polymerized material varies strongly with the feed ratio $x$.   
As  discussed in section \ref{sec:strict_alt}, in the strongly negative coupling limit the fraction of polymerized material reaches a maximum value $f^{\text{max}}$ that depends on the stoichiometric ratio $ \alpha $  and activation free energy $a$ as given by Eq.~\eqref{eq:fmax}. In  Fig.~ \ref{fig:expfrac}, we have also included the theoretical prediction of $f^{\text{max}}$.  We find  excellent agreement between the experimental and theoretical  trends for large activation free energies $ 4<a<8$.  
 These results confirm our hypothesis that  predominant configuration of assemblies are alternating  copolymers and   homopolymerization  is entirely suppressed due to the repulsive electrostatic interactions between identical species. Another important conclusion  that can be drawn is that the  supramolecular copolymerization is highly cooperative and takes place via a nucleation-elongation mechanism.\cite{Ahlers2}    We believe these interpretations to be valid  even though our model only takes into account nearest neighbor interactions.  The electrostatic interactions in the solution are rather short-ranged due to the screened nature of electrostatic interactions,$\lambda_D\approx 2 \ell$. Moreover, the effect of electrostatic interactions with monomers beyond the adjacent neighbors  could be captured by a renormalization of the interspecies binding free energy.~\cite{Netz}

\section{ Discussion and concluding remarks}
\label{sec:conclusion}

We have examined the supramolecular self-assembly behavior of a two-component system under the conditions that interspecies binding is favored over binding between identical species. Our theoretical framework is based on the  self-assembled nearest neighbor Ising model with a negative (anti-ferromagnetic) coupling constant. In the strongly negative coupling limit, {\it i.e.}, $J \ll -1$, our model reduces to that of strictly alternating copolymers. 
The polymerization is maximal for equal volume fraction of the two species ($\alpha=1$). For unequal volume fractions,  the maximum fraction of polymerized material, $f^{\text{max}}$,  is smaller than unity because an excess of the more abundant species remains in monomeric form. The $f^{\text{max}}$ and  the maximum mean length of copolymers not only depend on the stoichiometric ratio $\alpha$ but also on the activation free energies. Hence,  at a fixed overall monomer volume fraction, the mean degree of polymerization  can be controlled by changing the stoichiometric ratio. We have also obtained the functional dependence of the critical concentration, $\Phi^*$, on the stoichiometric ratio in this limit. The critical concentration of copolymers with sufficiently negative $J$  scales as $\Phi^* \sim \exp(2J)$ at stoichiometric ratios close to one  and it is strikingly smaller than the critical concentration of either of the homopolymers. The stoichiometric  dependence of the  critical concentration  of  predominantly alternating copolymers  is notably different from the case of self-assembly predominated by blocky  ordering  $J\gg 1$. In the latter case,  the critical concentration is bounded by the critical concentration of the two species and monotonically changes from one to the other $\Phi^*_B < \Phi^*  < \Phi^*_A$.~\cite{Sara}

Investigating the copolymerization behavior for a finite and sufficiently negative coupling constant, we find that the numerically obtained critical concentration shows a good agreement with our analytical results in the strong coupling limit.  Moreover, the copolymerization behavior up to moderate volume fractions is similar to that of strictly alternating copolymer configurations.  However, at  larger volume fractions well beyond $\Phi^*$, the copolymers lose their strict alternating order and   
 copolymer  and homopolymer configurations with defects become predominant, even for the case of equal volume fractions. 
  We have only presented results for the stoichiometric ratios   $\alpha \leq 1$. However, `all situations' are accounted for  as   our model is symmetric with respect to the particle species. 

 Our theoretical results shed light on the experimental findings  of oppositely charged comonomers that  self-assemble  into linear aggregates in aqueous solutions.~\cite{Besenius,Ahlers2}  Especially, they rationalize the dependence of fraction of polymerized material in the saturation limit  on the stoichiometric ratio. We find a very good agreement between our theoretical predictions and  the experimental results. Even though our approach is restricted to nearest neighbour interactions, our model provides a  good  description of the experimental trends due to strong self-screening of electrostatic interactions in alternating copolymers with complementary charges. We note that our results are relevant for the description of self-assembly behavior of any two-component system with  strongly favorable interspecies binding, as long as the interactions between the identical species are much weaker or repulsive. Other examples include chelating supramolecular polymers and chiral amplification in supramolecular polymers.

\begin{acknowledgments}
\noindent{We gratefully acknowledge the financial support from the German Research Foundation (http://www.dfg.de) within SFB TRR 146 (http://trr146.de)).} \\

\end{acknowledgments}

\noindent \textbf{\large{List of symbols}}\\
\begin{itemize}
\item $b_{ij} $: the magnitude of  the bonded interaction free energies  between two monomers of type $i\in\{A,B\}$ and $j\in\{A,B\}$, in units of the thermal energy $k_B T$;
\item $a_i $: the activation free energy  of species   $i\in\{A,B\}$ in units of $k_B T$;
 \item $Z_{N}$: the partition function of an assembly of length $N$;
\item $\rho(N)$: the number density of assemblies with degree of polymerization $N$;
\item $\overline{N} $: the number-averaged degree  of polymerization  including the monomers;
\item $\overline{N}_p $: the number-averaged degree  of polymerization excluding the monomers;
\item $J\equiv\frac{1}{4}(b_{AA}-2b_{AB}+b_{BB})$: the effective coupling constant in the Ising model;
 \item  $H\equiv \frac{1}{2}[(b_{AA}-b_{BB})+(\mu_A-\mu_B)]$: the magnetic field in the Ising model;
 \item $\bar{b}\equiv\frac{1}{4}(b_{AA}+b_{BB}+2b_{AB})$: the average binding free energy;
 \item $\mu_i$: the chemical potential of species $i\in\{A,B\}$;
 \item $\bar{\mu} \equiv 1/2(\mu_A+\mu_B)$: the average chemical potential;
 \item $\Delta \mu \equiv 1/2(\mu_A-\mu_B)$:  the difference in chemical potentials;
 \item $\lambda_\pm$: the eigenvalues of transfer matrix of Ising model;
\item $\Lambda_\pm \equiv \lambda_\pm \exp(\bar{b}+\bar{\mu})$: the effective fugacities of the bidisperse system;
\item $\Phi_i$: volume fraction of  molecules of species $i\in\{A,B\}$;
\item $\alpha=\Phi_A/ \Phi_B$: the ratio of volume fraction of the two species, the so-called stoichiometric ratio;
\item $f$: the mean fraction of polymerized material;
\item $f_i$: mean fraction of  homopolymers composed of monomers of type $i\in\{A,B\}$;
\item $f_{AB}$: mean fraction of strictly alternating copolymers composed of equal concentration of $A$ and $B$ monomers;
\item $\Phi_i^*= \exp(-b_i+a_i)$:  the critical volume fraction associated with species $i$, demarcating the transition from minimal assembly to assembly-predominated regime;
\item $\Phi_{AB}^*= 2 \exp(-b_{AB}+a)$: the critical volume fraction of alternating copolymers composed of equal concentrations of $A$ and $B$ species;
\item $\Phi^*(\alpha)$: the critical volume fraction of a bidisperse system at stoichiometric ratio $\alpha$;
\item $\phi^m_i$: the volume fraction of free monomers of species $i$;
 \item $\xi_0\equiv \exp(-2J)/2$: the correlation length of an antiferromagnetic chain in the limit $J \ll -1$ ; 
\item $f_ {\text{alt}}(N)$: the fraction of copolymers of size $N$ with perfect alternating  order relative to all the assemblies of the same length;
 \item $f_ {\text{B-homo}}(N)$: the fraction of homopolymers  consisting of  $B$  species  with size $N$ relative to all the assemblies of the same length;
\item $f_ {\text{alt}} ^{\text{tot}}$: the total fraction of alternating copolymers of any length;
\item  $f_ {\text{B-homo}} ^{\text{tot}}$: the  total fraction of $B$-homopolymers of any length;
\item $q$: the effective charge valency per monomer;
 \item  $b_{\text{el}}$: the magnitude of electrostatic interactions between two neighboring charged monomers;
  \item $\lambda_B \equiv e^2 \big/ (4 \pi \epsilon k_BT)$: the Bjerrum length in which  $e$ is the elementary charge and $\epsilon$ is the dielectric constant of the solvent;
   \item $\lambda_D$: the Debye screening length;
    \item $\ell$: the average spatial distance between two monomers in a copolymer; 
\end{itemize}
 \bibliography{aipsamp}

\end{document}